\documentclass{aastex631}
\usepackage{amsmath}
\usepackage{booktabs}
\usepackage{chngpage}
\usepackage{multirow}
\usepackage{subfigure}
\usepackage{threeparttable} 
\usepackage{float}
\usepackage{hyperref}
\usepackage[section]{placeins} 
\usepackage[figuresright]{rotating}

\usepackage{color}
\usepackage{longtable}

\shorttitle{Discern Misclassified flat-spectrum radio quasars from low-frequency peaked BL Lacertae objects }
\shortauthors{Liang et al.}

\begin{document}

\title{Discern Misclassified flat-spectrum radio quasars from low-frequency peaked BL Lacertae objects}

\correspondingauthor{Y.G. Zheng}

\email{$^\ast$ynzyg@ynu.edu.cn}
\email{$^{\dagger}$yangwg@ynnu.edu.cn}
\email{$^{\dagger\dagger}$kangshiju@alumni.hust.edu.cn}

\author[0009-0004-4978-0404]{S. Liang}
\affiliation{Department of Physics, Yunnan Normal University, Kunming, Yunnan, 650092, People's Republic of China}

\author{W.G. Yang$^\dagger$}
\affiliation{Department of Physics, Yunnan Normal University, Kunming, Yunnan, 650092, People's Republic of China}

\author[0000-0003-0170-9065]{Y.G. Zheng$^\ast$}
\affiliation{Department of Physics, Yunnan Normal University, Kunming, Yunnan, 650092, People's Republic of China}

\author[0000-0002-9071-5469]{S.J. Kang$^{\dagger\dagger}$}
\affiliation{School of Physics and Electrical Engineering, Liupanshui Normal University, Liupanshui, Guizhou, 553004, People's Republic of China}

\begin{abstract}
A sample of 312 low-frequency peaked BL Lacertae objects (LBLs) and 694 flat spectrum radio quasars (FSRQs) with the parameters both redshift and $\gamma$-ray photon spectral index ($\Gamma _\gamma$) is compiled from the active galactic nuclei (AGNs) Catalog Data Release 2 (4LAC-DR2) from Fermi-LAT. The multi-wavelength data of the sample sources are downloaded from the Space Science Data Center (SSDC), and then match the corresponding gamma-ray data from 4FGL-DR2. The synchrotron radiation peak frequency and Compton dominance (CD) parameters of the sources are obtained by using a log-parabolic to fit the average-state multi-wavelength spectral energy distribution. A support vector machine (SVM) in the $\log L_\gamma$-$\Gamma_\gamma$ frame is utilized to delineate the optimal boundary between FSRQs and LBLs sources. The 1$\sigma$ position of the Gaussian fitting on the histograms of the $\Gamma_\gamma$, $\log \nu^{syn}_{peak}$, and CD parameter distributions are also introduced. In the criterion, 25 FSRQ candidates are selected from LBL sample sources. 
The optical spectral identification result confirms that 8 out of 13 candidate sources available with the optical spectral data exhibit the relationship of $EW > 5 \mathring{\mathrm{A}}$.

\end{abstract}


\section{Introduction} \label{sec:intro}
Blazars are a special subclass of active galactic nuclei (AGNs) characterized by their relativistic jets being oriented almost directly toward Earth 
(e.g., \citealt{1986ApJ...310..317G}; \citealt{wills1992survey}; \citealt{1995PASP..107..803U}; \citealt{fan1996properties}; \citealt{1998A&AS..132...83B}; \citealt{2002A&A...390..431R}; \citealt{2006MNRAS.371.1243H}; \citealt{2007ApJ...669...96W}; \citealt{2024ApJ...962..122K}; \citealt{2024ApJ...967..104Z}). This unique orientation results in distinctive observational properties, such as extreme variability, high polarization, and a broad multiwavelength spectral energy distribution (SED). Typically, the SED of a blazar exhibits a double-bump structure in the $log \nu -log\nu f_\nu$ frame (e.g., \citealt{2009ApJ...692...32D}; \citealt{2018linspectral}; \citealt{2018RAA....18...56K}; \citealt{2020MNRAS.491.2771C}; \citealt{2022MNRAS.515.2215C}). Within the lepton-origin model, the low-energy bump is attributed to synchrotron emission, while the high-energy gamma-ray bump is explained by processes such as synchrotron self-Compton (SSC) and external Compton (EC) (e.g., \citealt{1992ApJ...397L...5M}; \citealt{1997A&A...320...19M}; \citealt{1998MNRAS.299..433F}; \citealt{2018RAA....18...56K}; \citealt{2019ApJ...873....7Z}; \citealt{xiong2020multicolor}).

From an observational point of view, blazars are traditionally divided into two main subclasses based on their optical emission line properties: flat-spectrum radio quasars (FSRQs), which exhibit strong broad emission lines, and BL Lacertae objects (BL Lacs), which have weak or absent emission lines.
BL Lac objects can be classified into distinct categories according to the positions of their SED peaks. Low-frequency peaked BL Lac objects (LBLs) are characterized by their low-energy peaks situated in the submillimeter to infrared spectrum, while their high-energy peaks reside within the GeV range. Intermediate-frequency peaked BL Lac objects (IBLs) display low-energy peaks in the optical domain and high-energy peaks around 10 GeV. High-frequency peaked BL Lac objects (HBLs) feature low-energy peaks extending from the ultraviolet to the X-ray band, with their high-energy peaks reaching into the TeV range and beyond (e.g., 
\citealt{1995ApJ...444..567P}; \citealt{2010A&A...512A..74M}).

Despite this classification, ambiguities remain. The equivalent width EW-based criterion (FSRQs have EW $\geqslant5\mathring{\mathrm{A}}$ and BL Lacs have EW $<5\mathring{\mathrm{A}}$) (e.g., \citealt{1995PASP..107..803U}; \citealt{1996MNRAS.281..425M}; \citealt{2003ApJ...588..128P}) has limitations, as changes in the EW, with values exceeding the 5$\mathring{\mathrm{A}}$ threshold, have been observed in the low jet activity state (\citealt{1995PASP..107..803U}; \citealt{2000MNRAS.311..485C}).
These effects were initially discussed by \cite{2012RAA....12..359F}, who anticipated that Doppler-enhanced continuum in FSRQs might obscure their spectral lines and that non-thermal radiation could swamp the broad emission lines, leading to misclassification. This idea was later explored and expanded by \cite{2021ApJS..253...46P} and \cite{2022MNRAS.515.2215C}, whose works are conceptually aligned with Foschini's earlier findings.

To address these limitations, \cite{ghisellini2011transition} proposed a physically motivated classification based on the luminosity of the broad-line region ($L_{BLR}$). In this scheme, FSRQs are defined as sources with a luminous BLR $L_{BLR}$ $\gtrsim$ $10^{-3}$, indicating radiatively efficient accretion ($L_{disk}$ $\gtrsim$ $L_{Edd}$). However, this approach requires knowledge of the source's redshift and black hole mass, which can complicate its application (\citealt{2017MNRAS.469..255G}). Furthermore, studies such as  \cite{2012MNRAS.420.2899G}  and \cite{linford2012gamma} have provided evidence that LBLs and FSRQs may originate from the same population, with some LBLs exhibiting characteristics of both subclasses. \cite{2019ApJ...879..107F} further demonstrated that the jet power distribution of LBLs reveals two distinct groups: one resembling FSRQs and the other resembling BL Lacs.

In this study, we compiled a sample of 694 FSRQs and 312 LBLs from the second release of the Fermi-LAT active galactic nuclei catalog (4LAC\text{--}DR2), complemented by multi-band data from the Space Science Data Center (SSDC). Using a log-parabolic model, we fitted the average-state multiwavelength SEDs of these sources to derive key physical parameters such as the effective spectral index, synchrotron peak frequency, and Compton dominance (CD). These parameters offer valuable insights into the intrinsic differences between LBLs and FSRQs. We employed a support vector machine (SVM) in the $\log L_\gamma$-$\Gamma_\gamma$ frame to delineate an optimal boundary between the two subclasses.

The paper is structured as follows:
In Section~\ref{sec:sample}, we introduce the sample selection and data compilation.
In Section~\ref{sec:model}, we describe the modeling of the SED and the derivation of physical parameters.
Section~\ref{sec:scr sour} outlines the methods used to filter the sources and present the results.
Section~\ref{sec:spectrum identification} discusses the spectral fitting and validation procedures. Finally, our conclusions are summarized in Section~\ref{sec:summary}. 
Throughout this work, we assume that the Hubble constant is $H_{0} = 75$ km s$^{-1}$ Mpc$^{-1}$, the dimensionless cosmological constant $\Omega_{\rm A} = 0.73$, the matter energy density $\Omega _M$=0.27, the radiation energy density $\Omega_R$=0 and the deceleration factor $q_0=0.5$.

\section{The sample description}     \label{sec:sample}

Launched by NASA on June 11, 2008, the Fermi Large Area Telescope (Fermi-LAT) covers an energy range of 50 MeV–1 TeV (\citealt{2020ApJS..247...33A}).
Our gamma-ray sample was obtained from the Fourth Fermi-LAT Source Catalog (4FGL), which contains 5064 gamma-ray sources. Using 4FGL as a reference, the Fourth Catalog of AGN Data (4LAC) identifies 694 FSRQs, 312 LBLs, 287 IBLs, 663 HBLs, and 1312 blazar candidates of uncertain type (BCUs). For our study, we selected data from the Second Data Release of 4LAC (4LAC-DR2\footnote{\scriptsize{\url{https://fermi.gsfc.nasa.gov/ssc/data/access/lat/4LACDR2/}}}) (\citealt{2020arXiv201008406L}), which compiles ten years of sky measurements by the Fermi Gamma-ray Space Telescope\footnote{\scriptsize{\url{https://tools.ssdc.asi.it/SED/}}}. This updated catalog includes 3131 sources in high Galactic latitudes ($| b |>  10^\circ$) and 380 sources in low Galactic latitudes ($| b |\leqslant 10^\circ$), totaling 3511 objects (\citealt{2020ApJ...892..105A}).
The 4LAC-DR2 catalog provides detailed information for each source, including classifications, redshifts, synchrotron radiation peak frequencies (frequency at the maximum synchrotron radiation flux), synchrotron radiation peak fluxes ($\nu f_\nu $), and corresponding source names in Very Long Baseline Interferometry (VLBI) observations (e.g.,\citealt{2010ApJ...716...30A}; \citealt{2019MNRAS.482.3023P}; \citealt{2020ApJ...892..105A}). To analyze these sources, we downloaded multi-band data from the SSDC\footnote{\scriptsize{\url{ https://tools.ssdc.asi.it/}}} and matched data from the seven gamma-ray energy bands of the LAT 10-year Source Catalog (4FGL-DR2\footnote{\scriptsize{\url{https://fermi.gsfc.nasa.gov/ssc/data/access/lat/10yr_catalog/}}})(\citealt{2020arXiv201008406L}). These bands cover the following energy ranges: Low-energy gamma rays (30-100 MeV), High-energy gamma rays (100 MeV-1 GeV), General gamma rays (1-3 GeV), Medium-energy gamma rays (3-10 GeV), High-energy gamma rays (10-30 GeV), Very high-energy gamma rays (30-100 GeV), Ultra-high-energy gamma rays ($>$100 GeV). These observations enable studies of cosmic rays, dark matter, and cosmic background radiation. To ensure the completeness of the data, we have also matched the Fermi 7 energy band data for these 1007 sources from the Fourth Fermi Gamma-ray Source Catalog (4FGL-DR2), including the radiation flux values and their associated uncertainties at frequencies of approximately 22.26, 22.68, 23.2, 23.68, 24.2, 24.68, and 25.6. These detailed measurements provide a foundation for further investigation into the physical properties and classifications of blazars.

\section{The Compton Dominance Parameter}\label{sec:model}
In the random acceleration process, considering only energy loss via radiative cooling, we can obtain the logarithmic parabolic electron energy distribution (e.g., \citealt{2004A&A...413..489M}; \citealt{2006A&A...448..861M}; \citealt{2021RAA....21....8Z}; \citealt{2021ApJ...915...59Z}):

\begin{equation}{\label{equ:1}}
N_e(\gamma)=N_0\left(\frac{\gamma}{\gamma_0}\right)^{-s-r\log\left(\frac{\gamma}{\gamma_0}\right)},
\end{equation}
where, $N_0$ is the normalization coefficient, $\gamma_0$ is the Lorentz factor of the initial electron, $s$ denotes the spectral index, and $r$ represents the spectral curvature (\citealt{2024A&A...685A.140R}). We apply this function form to the radiation flux, then the radiation flux at different radiation frequencies $\nu$ can be expressed as ( e.g., \citealt{1962SvA.....6..317K}; \citealt{2009A&A...504..821P}; \citealt{2016ApJ...819..156B}):

\begin{equation}{\label{equ:2}}
f(\nu)=k(\frac \nu{\nu_0})^{-\alpha_0-\beta \log{(\frac \nu{\nu_0})}},
\end{equation}
where $\nu _0$ is the reference frequency, $\alpha _0$ equals the local power-law photon index at $\nu _0 $, and $\beta $ is the curvature of the energy spectral bump. In the space coordinate system $\log \nu-\log (\nu f_\nu)$, the above equation can be written as (\citealt{2021ApJ...915...59Z}):
\begin{equation}{\label{equ:3}}
\begin{aligned}
\log\left(\nu f_{\nu}\right)& =\log{[k(\frac{\nu}{\nu_{0}})^{-\alpha_{0}-\beta\log{(\frac{\nu}{\nu_{0}})}}]}+\log{\nu} \\
&=-\beta(\log\nu)^{2}+(1-\alpha_{0}+2\beta\log \nu_{0})\mathrm{log} \nu +(\log k+\alpha_0\mathrm{log} \nu_0-\beta(\log\nu_0)^2),
\end{aligned}
\end{equation}
where, we set $A=-\beta, B=1-\alpha_0+2\beta\mathrm{log} \nu_0, C=\log k+\alpha_0\mathrm{log} \nu_0-\beta(\log\nu_0)^2$, then the above formula can be writen as:

\begin{equation}{\label{equ:4}}
\log \nu f_\nu=A(\log \nu)^2+B\log \nu+C.\end{equation}

The above formula is a logarithmic parabola model. We can get the parameters of the photon spectrum by replacing the power law spectrum of the flare non-thermal radiation. Generally, the non-thermal radiation of the flare is expressed as (\citealt{2016ApJS..226...20F}):

\begin{equation}{\label{equ:5}}
f_{\nu}\propto\nu^{-\alpha},
\end{equation}
where, $f_\nu$ is the flow at frequency $\nu $, $\alpha$ is the spectral index, in the space coordinates of $\log\nu-\log(\nu f_\nu) $, equation (\ref{equ:4}) can be converted to:

\begin{equation}{\label{equ:6}}
\log\nu f_\nu=k+(1-\alpha)\mathrm{log~}\nu,
\end{equation}
 where, $k$ is a constant.  Combining equation (\ref{equ:4}) and equation (\ref{equ:6}) the spectral slope rate at $\nu = \nu_1$ can be obtained:

\begin{equation}{\label{equ:7}}
\left[\frac{d\left(\log \nu f_{\nu}\right)}{d\left(\log \nu \right)}\right]_{\nu=\nu_{1}}=2A\log\nu_{1}+B=1-\alpha_{\nu_{1}},
\end{equation}
the peak frequency of energy spectrum is:

\begin{equation}{\label{equ:8}}
\nu_{\mathrm{peak}}=10^{-\frac B{2A}},
\end{equation}
the peak flow of the energy spectrum is:

\begin{equation}{\label{equ:9}}
f_{\mathrm{peak}}=k\Bigg(\frac{\nu_{\mathrm{peak}}}{\nu_0}\Bigg)^{-\alpha_0-\beta\log\Bigg(\frac{\nu_{\mathrm{peak}}}{\nu_0}\Bigg)},
\end{equation}
the peak luminosity is (\citealt{2016ApJS..226...20F}):

\begin{equation}{\label{equ:10}}
L_\mathrm{peak}=4\pi d_L^2\nu_\mathrm{peak}f_\mathrm{peak} ,
\end{equation}
where $d_L\left(=(1+z)\cdot\frac c{H_0}\cdot\int_1^{1+z}\frac1{\sqrt{\Omega_Mx^3+1-\Omega_M}}dx\right)$(\citealt{2007NJPh....9..445C}; \citealt{2016ApJS..226...20F}) is luminosity distance and $f_\nu$ is the K-corrected flux density at the corresponding frequency $\nu$. The Compton dominance parameter is defined as the ratio of inverse Compton scattering peak luminosity to synchrotron radiation peak luminosity (\citealt{2021ApJS..253...46P}), so we have:

\begin{equation}{\label{equ:11}}
CD=4\pi d_L^2\nu_\text{ic,peak}f_\text{ic,peak}/4\pi d_L^2\nu_\text{syn,peak}f_\text{syn,peak},
\end{equation}
since the redshift above and below the equation can be eliminated, it can also be viewed as the ratio of the flow of high energy peaks to low energy peaks, i.e:

\begin{equation}{\label{equ:12}}
CD=\frac{\nu_{ic}f_{\nu,ic}}{\nu_{syn}f_{\nu,syn}}.
\end{equation}

According to the definition of Compton's dominant parameter, \cite{2013ApJ...763..134F} derived it in detail. Here, we will make a brief elaboration. Here, CD can be expressed as:
\begin{equation}{\label{equ:13}}
CD\equiv\frac{\max\left[L_{\mathrm{pk}}^{\mathrm{EC}},L_{\mathrm{pk}}^{\mathrm{SSC}}\right]}{L_{\mathrm{pk}}^{\mathrm{sy}}},
\end{equation}
here, $L_{\mathrm{pk}}^{\mathrm{EC}}$ is the peak luminosity of the EC process,  
$L_{\mathrm{pk}}^{\mathrm{SSC}}$ is the peak luminosity of the SSC process, and $L_{\mathrm{pk}}^{\mathrm{sy}}$ is the peak luminosity of the synchrotron radiation process.

According to equation (\ref{equ:13}), the peak luminosity from Thomson scattering an external isotropic radiation field in the Dirac $\delta$-function is (\citealt{2002ApJ...575..667D}):
\begin{equation}{\label{equ:14}}
L_{\mathrm{pk}}^{\mathrm{EC}}=\delta_{\mathrm{D}}^{6}c\sigma_{\mathrm{T}}u_{\mathrm{ext}}Q_{0}t_{\mathrm{csc}}\begin{cases}\gamma_{c}^{3-q}&\gamma_{1}<\gamma_{c}\\\gamma_{c}\gamma_{1}^{2-q}&\gamma_{c}<\gamma_{1}\end{cases}.\end{equation}
where, $\delta_{\mathrm{D}}$ is Doppler factor, $Q_0$ is the normalization constant of the electron injection distribution, $\sigma_{\mathrm{T}}$ is the Thomson scattering cross-section, $c$ is the speed of light in vacuum, $u_{\mathrm{ext}}$ is the energy density of the external photon field, and $t_{\mathrm{csc}}$ represents an energy-independent escape timescale.
The SSC luminosity at the peak in the Thomson regime can be approximated by \cite{2008ApJ...686..181F}:

\begin{equation}{\label{equ:15}}
L_{\mathrm{pk}}^{\mathrm{SSC}}=\frac{\delta_\mathrm{D}^4}{3\pi R_b^{\prime2}}c\sigma_\mathrm{T}^2u_B^{\prime}(Q_0t_{\mathrm{esc}})^2\begin{cases}\gamma_c^{6-2q}&\gamma_1<\gamma_c\\\gamma_c^2\gamma_1^{4-2q}&\gamma_c<\gamma_1\end{cases}.\end{equation}
 $R_b^{\prime}$ is the radius of the emission region in the jet in the comoving frame, and $u_B^{\prime}$ is the magnetic field energy density within the jet's emission region.
The synchrotron radiation peak flow density is:
\begin{equation}{\label{equ:16}}
u_{\mathrm{sy,pk}}^{\prime}=\frac{R_b^{\prime}}{c}\frac{L_{\mathrm{pk}}^{\prime\mathrm{sy}}}{4\pi R_b^{\prime3}/3}=\frac{u_B^{\prime}\sigma_\mathrm{T}}{2\pi R_b^{\prime2}}Q_0t_{\mathrm{esc}}\begin{cases}\gamma_c^{3-q}&\gamma_1<\gamma_c\\\gamma_c\gamma_1^{2-q}&\gamma_c<\gamma_1\end{cases}.\end{equation}

So, combine with equation (\ref{equ:14})(\ref{equ:15})(\ref{equ:16}), the CD can also be represented as:
\begin{equation}{\label{equ:17}}
CD\approx\frac{\max\left[\delta_\mathrm{D}^2u_\mathrm{ext},u_\mathrm{sy,pk}^{\prime}\right]}{u_B^{\prime}}.\end{equation}

The accretion disk provides an isotropic distribution of the soft photon field and the energy density of the outer photon field,

\begin{equation}{\label{equ:18}}
u_\mathrm{ext}=u_\mathrm{d}+u_\mathrm{BLR}+u_\mathrm{MT}.\end{equation}

Energy density of photon field outside the accretion disk:

\begin{equation}{\label{equ:19}}
u_d\approx\frac{\Gamma^2L_d^{\prime}\tau_d}{3\pi R_d^{\prime2}c}[erg cm^{-3}],\end{equation}
here, $\Gamma$  represents the bulk Lorentz factor of the jet, $L^{\prime}_d$ denotes the luminosity of the accretion disk surrounding  the central supermassive black hole, $\tau_d$ is the optical depth of the photons from the accretion disk scattered by the surrounding medium, and $R^{\prime}_d$ represents the squared radius of the region where the accretion disk photons are scattered by the surrounding medium. The broad-line region and the dust ring contribute to the external photon field, leading to differences in their respective energy densities.

\begin{equation}{\label{equ:20}}u_{\mathrm{BLR}}(r^{\prime})=\frac{\tau_{BLR}\Gamma^2L_d^{\prime}}{3\pi R_{BLR}^2c[1+(r^{\prime}/R_{BLR}^{\prime})\beta_{BLR}]} [erg cm^{-3}],
\end{equation}

\begin{equation}{\label{equ:21}}
u_{MT}(r^{\prime})=\frac{\tau_{MT}\Gamma^2L_d^{\prime}}{3\pi R_{MT}^2c[1+(r^{\prime}/R_{MT}^{\prime})\beta_{MT}]} [erg cm^{-3}],\end{equation}
In the formulas (\ref{equ:20}), (\ref{equ:21}), $r^{\prime}$ represents the distance between the energy dissipation region and the central black hole horizon in the jet moving coordinate system,  $L^{\prime}_d$ denotes the luminosity of the accretion disk around the central supermassive black hole, $\tau_{BLR}$ and $\tau_{MT}$ are the fractional proportions of the photons entering the broad line region and the dust ring, respectively, relative to the disk luminosity, $R^{\prime}_{BLR}$ and $R^{\prime}_{MT}$  indicate  characteristic distances of the broad line region and the dust ring, while $\beta_{BLR}$  and $\beta_{MT}$ are the radiation density profile coefficients.
The SSC model has been widely used to fit the multiwavelength SED of BL Lac objects (e.g., \citealt{2004ApJ...601..151K}; \citealt{zhang2012radiation}). In contrast, a combination of the SSC and EC models is often applied to describe FSRQs more accurately (e.g., \citealt{2002ApJ...564...92C}; \citealt{2011MNRAS.414.2674G}; \citealt{2014MNRAS.439.2933Y}). Consequently, the CD of FSRQs is generally higher than that of BL Lacs.                 

\begin{figure*}[htbp]
	\centering
 \includegraphics[width=1\linewidth]{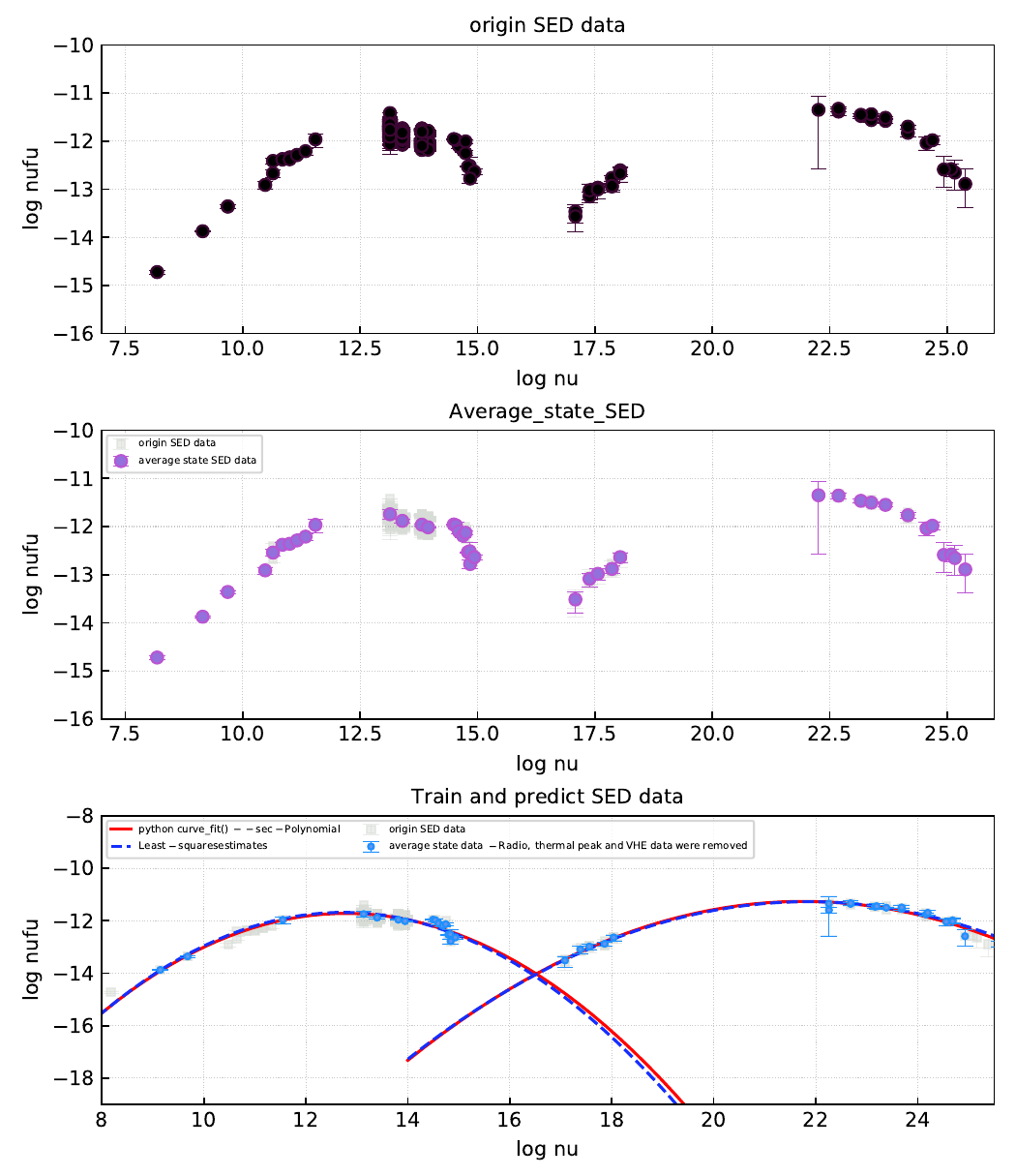}
 
	\caption{Multi-band data processing fitting process in the $\log\nu-\log\nu f_\nu$ space. The black points are the multi-band data points downloaded from the SSDC and 4FGL-DR2. the second figure We take the arithmetic average of multiple data points at the same radiation frequency to get. The purple points are the points obtained after taking the arithmetic average, and the gray points are the points that do not participate in the fitting after taking the arithmetic average.}
	\label{fig:1}
\end{figure*}

\begin{figure*}[htbp]
  \centering
  \includegraphics[trim=0cm 6cm 0cm 0cm, clip, width=0.8\linewidth]{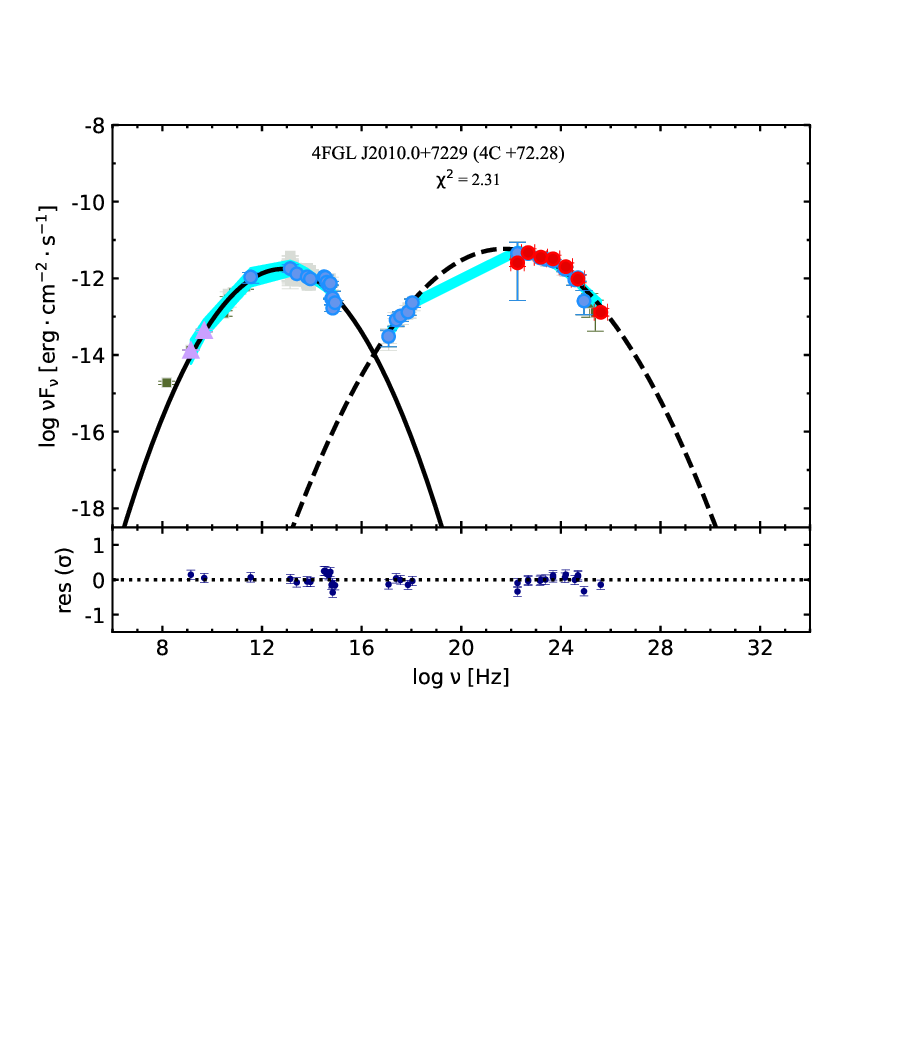}
 
  \caption{Energy spectrum distribution of Fermi blazars in the $log\nu-log\nu f_\nu$ space. The black solid line is the fitting curve of the synchrotron radiation peak, and the black dashed line is the fitting line of the IC scattering peak. The black dots are multi-band data points. The red data points are the 7 energy bands of Fermi, and the blue areas are the confidence ranges, and the small plot below shows how each data point deviates from the fitted curve.}
   \label{fig:2}
\end{figure*}

To enhance the accuracy of the energy spectrum distribution for each source, we applied several data processing techniques when fitting the blazar sources. These include:
For multiple data points at the same frequency, their arithmetic mean was calculated to mitigate the influence of repeated points on the fitting process.
Data points with energy uncertainties larger than their flux values were removed to exclude potential noise or interference.
Redshift values were retrieved from the NASA/IPAC Extragalactic Database (NED). If no redshift measurement was available in NED, the value was selected from the 4LAC-DR2 catalog.
All energy spectra are presented in the  $log\nu - log(\nu f_\nu)$  space, as illustrated in Figure \ref{fig:1}. The black points represent multi-band data obtained from the official Fermi website, including the seven energy bands of Fermi. After taking the arithmetic mean of data points with the same radiation frequency, the averaged points are shown in purple in Figure \ref{fig:1}, while gray points indicate those excluded from the fitting process. We fit the energy spectrum data using least squares and a logarithmic parabola, as displayed in the third panel of Figure \ref{fig:1}. The red solid line represents the logarithmic parabola fit, while the blue dashed line corresponds to the least squares fit. Our program automatically selects the optimal curve for the subsequent calculation of physical parameters (\citealt{2021ApJ...915...59Z}). The final fitted curve is presented in Figure \ref{fig:2}.

In the energy spectrum of the bimodal structure shown in the figures above, the peak of the first black solid line represents the synchrotron radiation generation. The black dotted line corresponds to the fitting curve for inverse Compton scattering. The smaller figure below illustrates the deviation of each data point from the fitting line. Using the fitting parameters, we calculated the peak frequency, radiation peak flux, and Compton dominance (CD) parameters for the synchrotron radiation and inverse Compton scattering.

\section{Screening the misclassified sources} \label{sec:scr sour}
By applying logarithmic parabola fitting, we determined the synchrotron radiation peak frequency for LBL and FSRQ sources, the inverse Compton scattering peak frequency, synchrotron radiation peak flux, inverse Compton scattering peak flux, and CD. Table \ref{tab:1} and Table \ref{tab:2} summarize these results.

\begin{table}[!ht]
    \centering
    \caption{Logarithmic parabola fitting results for LBLs}
    \label{tab:1}
    \begin{tabular}{ccccccccccc}
    \hline\hline
     $\rm 4FGL\;Name$&$\rm Source name$  & ${\rm Redshift}$ & $ \Gamma_\gamma$ & $log \nu_{peak}^{syn}$ & $log \nu_{peak}^{ic}$ & $log \nu {f_\nu} _{peak}^{syn}$ & $log \nu {f_\nu} _{peak}^{ic}$ & $CD$  & $L_\gamma$ \\
        \normalsize(1) & \normalsize(2) & \normalsize(3) &\normalsize(4) & \normalsize(5)   &\normalsize(6) &\normalsize(7) &\normalsize(8) &\normalsize(9) &\normalsize(10)\\
        \hline
      J0001.2-0747 & PMN J0001-0746 & 0.966  & 2.148  & 18.199  & ~ & -10.655  & ~ & ~ &   \\ 
         J0003.2+2207 & 2MASX J00032450+2204559 & 0.966  & 2.276  & 14.291  & 20.630  & -12.080  & -10.973  & 12.803  & 45.613   \\ 
         J0003.9-1149 & PMN J0004-1148 & 0.860  & 2.128  & 12.673  & 22.528  & -12.025  & -12.323  & 0.504  & 44.966   \\ 
         J0006.3-0620 & PKS 0003-066 & 0.347  & 2.171  & 12.433  & 19.712  & -10.900  & -12.133  & 0.059  & 43.916   \\ 
         J0008.0+4711 & MG4 J000800+4712 & 0.280  & 2.053  & 14.137  & 22.993  & -11.954  & -11.505  & 2.814  & 44.905   \\ 
         J0009.1+0628 & TXS 0006+061 & 1.563  & 2.117  & 13.499  & 23.214  & -11.872  & -12.066  & 0.640  & 45.847   \\ 
         J0013.1-3955 & PKS 0010-401 & 1.563  & 2.050  & 12.161  & 8.181  & -11.936  & -16.490  & 0.000  & 46.478   \\ 
         J0014.1+1910 & MG3 J001356+1910 & 0.477  & 2.264  & 13.829  & 22.979  & -11.813  & -11.978  & 0.685  & 44.661   \\ 
         J0019.6+2022 & PKS 0017+200 & 0.858  & 2.152  & 12.500  & 22.265  & -11.839  & -12.206  & 0.430  & 44.978   \\ 
         J0022.1-1854 & 1RXS J002209.2-185333 & 0.774  & 1.935  & 14.488  & ~ & -10.336  & ~ & ~ &   \\ 
         J0022.5+0608 & PKS 0019+058 & 2.860  & 2.170  & 13.132  & 22.646  & -11.782  & -11.432  & 2.239  & 47.165   \\ 
         J0023.9+1603 & 87GB 002122.5+154553 & 0.770  & 2.066  & 13.549  & 24.115  & -12.206  & -12.126  & 1.203  & 45.161   \\ 
         J0029.0-7044 & PKS 0026-710 & 0.770  & 2.224  & 14.201  & 23.773  & -11.945  & -10.576  & 23.383  & 45.557   \\ 
         J0032.4-2849 & PMN J0032-2849 & 0.324  & 2.297  & 13.828  & 20.722  & -11.810  & -11.771  & 1.094  & 43.861   \\ 	\hline

            \end{tabular}
            \tablecomments{Columns (1) is the 4FGL name of sources; Columns (2) is the associated source name; Columns (3) is the redshift; Columns (4) is the $\gamma$ -ray photon spectral index; Columns (5) is the Synchrotron radiation peak frequency; Columns (6) is the Inverse Compton scattering peak frequency; Columns (7) is the Synchrotron radiation peak flow; Columns (8) is the Inverse Compton scattering peak flow; Columns (9) is the Compton dominant parameter; Columns (9) is $\gamma$ -ray luminosity. (This table is available in its entirety in machine-readable form.)	}
\end{table}
	         	
\begin{table}[!ht]
    \centering
    \caption{Logarithmic parabola fitting results for FSRQs}
    \label{tab:2}
    \begin{tabular}{ccccccccccc}
    \hline\hline
     $\rm 4FGL\;Name$&$\rm Source name$  & ${\rm Redshift}$ & $ \Gamma_\gamma$ & $log \nu_{peak}^{syn}$ & $log \nu_{peak}^{ic}$ & $log \nu {f_\nu} _{peak}^{syn}$ & $log \nu {f_\nu} _{peak}^{ic}$ & $CD$  & $L_\gamma$ \\
        \normalsize(1) & \normalsize(2) & \normalsize(3) &\normalsize(4) & \normalsize(5)   &\normalsize(6) &\normalsize(7) &\normalsize(8) &\normalsize(9) &\normalsize(10)\\
        \hline
        J0001.5+2113 & TXS 2358+209 & 1.1060  & 2.6802  & 14.1457  & 20.7915  & -11.9120  & -9.8058  & 127.7019  & 47.0224   \\ 
         J0004.4-4737 & PKS 0002-478 & 1.1060  & 2.4150  & 13.8432  & 20.7686  & -11.4264  & -10.3176  & 12.8475  & 45.5259   \\ 
         J0005.9+3824 & S4 0003+38 & 1.1060  & 2.6668  & 12.6789  & 20.6292  & -11.2398  & -10.3552  & 7.6658  & 44.9021   \\
         J0010.6+2043 & TXS 0007+205 & 1.1060  & 2.3165  & 12.8816  & 20.2438  & -12.2573  & -11.8627  & 2.4810  & 45.1100   \\ 
         J0010.6-3025 & PKS 0008-307 & 1.1900  & 2.4291  & 13.0010  & 20.8813  & -12.1038  & -11.3588  & 5.5592  & 45.2328   \\ 
         J0011.4+0057 & RX J0011.5+0058 & 1.4909  & 2.3198  & 13.0223  & 21.1131  & -11.9192  & -11.2185  & 5.0198  & 45.5583   \\ 
         J0013.6-0424 & PKS 0011-046 & 1.0750  & 2.3587  & 12.8877  & 20.6222  & -12.1214  & -11.1338  & 9.7182  & 45.4163   \\ 
         J0016.2-0016 & S3 0013-00 & 1.5764  & 2.7265  & 12.9644  & 20.8392  & -12.2004  & -10.5822  & 41.5096  & 45.3982   \\ 
         J0016.5+1702 & GB6 J0015+1700 & 1.7090  & 2.6309  & 12.6801  & 20.7086  & -12.0956  & -10.9771  & 13.1343  & 45.6553   \\ 
         J0017.5-0514 & PMN J0017-0512 & 0.2270  & 2.5346  & 13.8541  & 20.7302  & -11.4951  & -9.8088  & 48.5560  & 43.9322   \\ 
         J0019.6+7327 & S5 0016+73 & 1.7810  & 2.5942  & 12.5681  & 20.6175  & -11.2670  & -9.9283  & 21.8147  & 46.9898   \\ 
         J0023.7+4457 & B3 0020+446 & 1.0620  & 2.4420  & 12.6660  & 21.3858  & -11.9694  & -11.5576  & 2.5807  & 45.3692   \\ 
         J0024.7+0349 & GB6 J0024+0349 & 0.5450  & 2.3887  & 12.5923  & 23.7352  & -11.9560  & -11.9588  & 0.9936  & 44.9352   \\ 
         J0025.2-2231 & PMN J0025-2228 & 0.8340  & 2.4007  & 14.0741  & 23.9843  & -12.1415  & -12.7665  & 0.2371  & 44.7952   \\ 	\hline

            \end{tabular}
            \tablecomments{Columns (1) is the 4FGL name of sources; Columns (2) is the associated source name; Columns (3) is the redshift; Columns (4) is the $\gamma$ -ray photon spectral index; Columns (5) is the Synchrotron radiation peak frequency; Columns (6) is the Inverse Compton scattering peak frequency; Columns (7) is the Synchrotron radiation peak flow; Columns (8) is the Inverse Compton scattering peak flow; Columns (9) is the Compton dominant parameter; Columns (10) is $\gamma$ -ray luminosity. (This table is available in its entirety in machine-readable form.)	}
\end{table}

\begin{figure*}[!ht]
	\centering
	\includegraphics[width=0.7\linewidth]{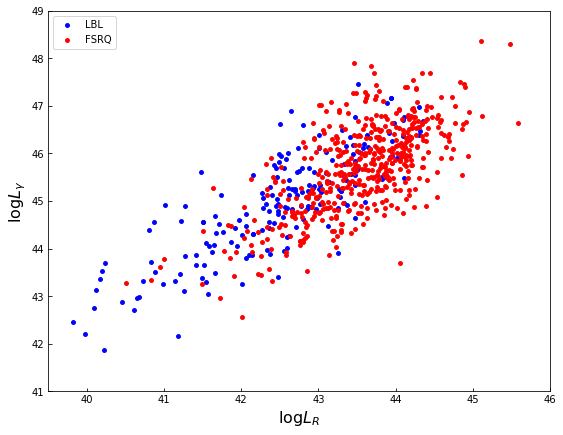}
	\caption{The relationship between $\log L_\gamma$ (y-axis) and $\log L_R$ (x-axis). The scatter plot shows the relationship between $\log L_\gamma$ and $\log L_R$ on a logarithmic scale. Blue points represent LBL sources, while red points represent FSRQ sources. A positive correlation is observed between $\log L_\gamma$
  and $\log L_R$ for both types, with FSRQ sources generally exhibiting higher values in both luminosities compared to LBL sources.}
	\label{fig:3}
\end{figure*}

There is a strong linear relationship between $L_\gamma$ and $L_R$ in Figure \ref{fig:3}, as confirmed by the regression analysis results shown in Table \ref{tab:3}. This finding aligns with the results of  \cite{osti_22092298}, which demonstrate that the energy spectrum of blazars can be effectively modeled using a logarithmic parabola model.

\begin{table}[!ht]
    \centering
    \tablenum{3}
    \caption{Results of the regression analysis.}
    \label{tab:3}
    \begin{tabular}{ccccccccc} 
    \hline\hline
        \multirow{2}{*}{Parameter A} & \multirow{2}{*}{Parameter B} & \multicolumn{3}{c}{LBLs} & \multicolumn{3}{c}{FSRQs} & \multirow{2}{*}{Notes} \\ \cline{3-8} 
        & & N & $\rho$ & $p$ & N & $\rho$ & $p$ &  \\ \hline
        \multirow{2}{*}{$\log L_R$} & \multirow{2}{*}{$\log L_\gamma$} & 273 & 0.893 & $<0.0001$ & 613 & 0.690 & $<0.0001$ & present work \\ 
        & & 118 & 0.840 & $<0.0001$ & 676 & 0.790 & $<0.0001$ & \citealt{2022ApJS..262...18Y} \\ \hline
    \end{tabular}
    \tablecomments{Column 1 and column 2 are two parameters applied to the correlation analysis. $N$ is the number of sources in the sample. $\rho$ is the correlation coefficient of the Spearman correlation test. $p$ is the statistical p-value.}
\end{table}

\begin{figure*}[htbp]
	\centering
	\includegraphics[width=0.7\linewidth]{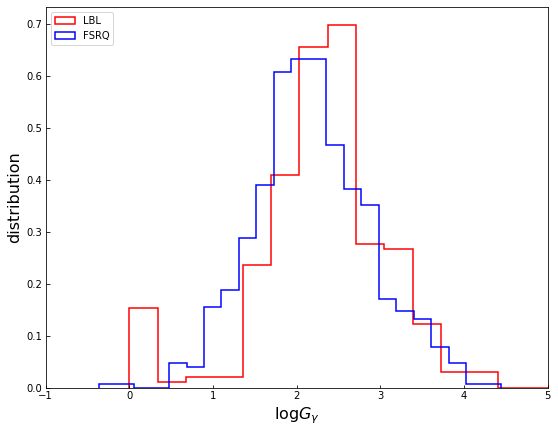}
	\caption{This figure illustrates the distribution of $G_\gamma$. The x-axis represents the logarithmic values of gamma-ray loudness, with the red curve representing the log-gamma ray loudness distribution for LBL and the blue curve representing the log-gamma ray loudness distribution for FSRQ. From the figure, it can be observed that the distribution ranges for both types of sources largely overlap, indicating their similarity in gamma-ray loudness.}
	\label{fig:4}
\end{figure*}

At the same time, we calculated the $\gamma$-ray loudness values using (\ref{equ:22}). The distribution indicates that the range for LBL and FSRQ is nearly identical. Therefore, based on the $G_\gamma$ distribution shown in Figure \ref{fig:4}, we infer that some sources may have been misclassified.
\begin{equation}{\label{equ:22}}
  G_\gamma=\frac{L_\gamma}{L_R}. 
\end{equation}

\begin{figure*}[htbp]
\centering
\includegraphics[width=0.618\linewidth]{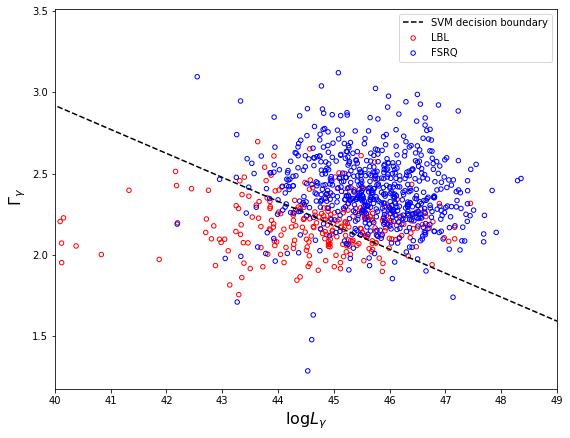}
\caption{The relationship between $\log L_\gamma$ and $\Gamma_\gamma$ based on a Support Vector Machine (SVM) classification model. Red points represent LBL, while blue points correspond to FSRQ. The dashed line indicates the SVM decision boundary, separating the two classes by $\Gamma_\gamma =- 0.147log L_\gamma+8.819 $. }
\label{fig:5}
\end{figure*}

Firstly, we perform a preliminary classification using a Support Vector Machine (SVM) to filter out potential misclassified sources in LBLs. SVM is a powerful supervised learning model designed for classification tasks. Its core principle is to identify an optimal hyperplane that maximally separates different classes by maximizing the margin—the distance between the support vectors (i.e., the closest data points) and the hyperplane. This margin optimization enhances the model’s generalization ability, enabling accurate predictions even with limited training samples.
 Using the support vector machine algorithm, we find the dividing line between FSRQ and LBL as follows: $\Gamma =- 0.147\log L_\gamma+8.819$ as shown in Figure \ref{fig:5}.  Based on this criterion, we selected 118 LBLs located above the black dotted line.

We then analyzed the distribution of the spectral indices of $\gamma$-ray photons for both types of objects. The boundary was set at the 1$\sigma$ value on the positive side of the FSRQ distribution curve, closest to the LBL, which corresponds to 2.267. Applying this criterion, we identified 48 LBLs with $\gamma$-ray photon spectral indices greater than 2.267, as shown in Figure \ref{fig:6} (a). Similarly, we selected 40 sources with $\log\nu _{peak}^{syn} < 13.597$, as illustrated in Figure \ref{fig:6} (b).  Finally, we used the CD parameter to determine the misclassified sources. A Gaussian fitting method was employed to determine the critical value, screening out 25 sources with CD values exceeding 0.776,  as detailed in Table \ref{tab:4}:
\begin{figure*}[htbp]
	\centering
	\subfigbottomskip=2pt
	\subfigcapskip=-5pt
	\begin{adjustwidth}{-0.0cm}{1cm}
		\subfigure[]{
	\includegraphics[width=0.5\linewidth]{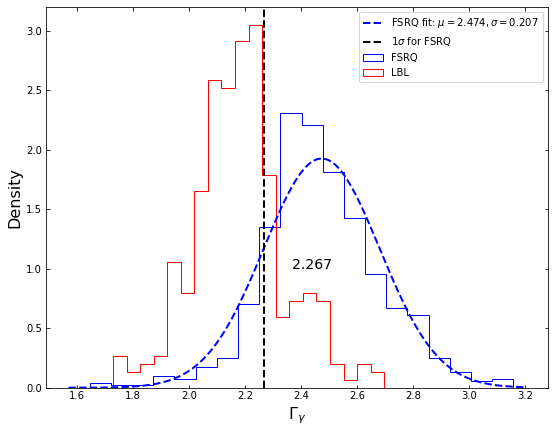}}\hspace{-1mm}
		\subfigure[]{
  \includegraphics[width=0.5\linewidth]{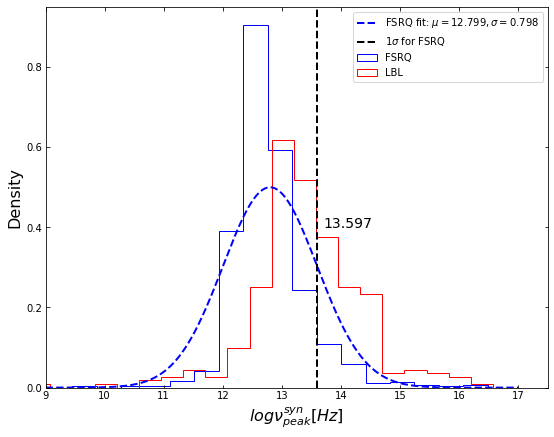}}\hspace{-1mm}\\
  \centering
		\subfigure[]{
  \includegraphics[width=0.6\linewidth]{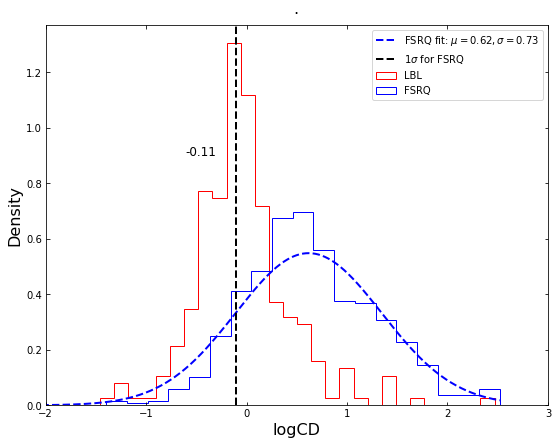}}\hspace{-1mm}  
	\end{adjustwidth}
	\caption{(a) the distribution of the $\gamma$ -ray photon spectral index; (b) is the distribution of $\log\nu _{peak}^{syn}$; (c) is the distribution of CD. Among them, the red solid line depicts the histogram of the LBL, while the blue solid line represents that of the FSRQ. The black dashed line indicates the value obtained through Gaussian fitting of the histogram of the FSRQ and within the $1\sigma$ confidence interval. This value is closer to one side of the histogram of the LBL. In the end, we drew the following dividing lines: $\Gamma_\gamma =2.267, \log\nu _{peak}^{syn} =13.597; CD=0.776$.}
	\label{fig:6}
\end{figure*}

\begin{table}[!ht]
    \centering
    \caption{The results of screening the misclassified sources}
    \label{tab:4}
    \begin{tabular}{ccccccccccc}
    \hline\hline
        4FGL name & source name & redshift & $\Gamma\gamma$ & $log \nu_{peak}^{syn}$ & $log \nu_{peak}^{ic}$ & $log \nu {f_\nu}_{peak}^{syn}$ & $log \nu {f_\nu}_{peak}^{ic}$ & CD & $\log L_\gamma$\\
        \normalsize(1) & \normalsize(2) & \normalsize(3) &\normalsize(4) & \normalsize(5)   &\normalsize(6) &\normalsize(7) &\normalsize(8) &\normalsize(9) &\normalsize(10) \\
        \hline
         J0348.6-1609 & PKS 0346-163 & 0.239  & 2.394  & 12.891  & 20.409  & -11.266  & -10.893  & 2.361  & 43.951   \\ 
         J0407.5+0741 & TXS 0404+075 & 1.133  & 2.286  & 12.659  & 21.948  & -11.639  & -11.712  & 0.845  & 45.390   \\ 
         J0438.9-4521 & PKS 0437-454 & 0.970  & 2.401  & 12.883  & 21.356  & -12.003  & -11.509  & 3.120  & 45.458   \\
         J0610.1-1848 & PMN J0610-1847 & 0.640  & 2.363  & 13.127  & 21.077  & -11.588  & -11.616  & 0.938  & 44.891   \\ 
         J0710.9+4733 & S4 0707+47 & 1.292  & 2.611  & 13.189  & 20.821  & -11.811  & -11.332  & 3.013  & 45.489   \\ 
         J0832.4+4912 & OJ 448 & 0.548  & 2.446  & 13.101  & 21.262  & -11.712  & -11.776  & 0.863  & 44.483   \\ 
         J0934.3+3926 & GB6 J0934+3926 & 0.236  & 2.540  & 12.700  & 21.629  & -12.045  & -11.875  & 1.480  & 43.761   \\ 
         J0941.9+2724 & GB6 J0941+2721 & 0.631  & 2.413  & 13.279  & 23.903  & -12.274  & -12.082  & 1.555  & 44.623   \\ 
         J1035.6+4409 & 7C 1032+4424 & 0.444  & 2.400  & 13.407  & 19.669  & -12.400  & -11.835  & 3.679  & 44.017   \\ 
         J1128.8+3757 & NVSS J112903+375655 & 4.090  & 2.316  & 13.571  & 22.115  & -12.253  & -11.930  & 2.102  & 46.891   \\ 
         J1135.1+3014 & CRATES J113514+301001 & 4.090  & 2.317  & 12.928  & 20.933  & -12.356  & -10.966  & 24.549  & 47.450   \\ 
         J1142.0+1548 & MG1 J114208+1547 & 0.299  & 2.336  & 12.917  & 22.540  & -12.033  & -11.865  & 1.474  & 44.299   \\ 
         J1148.6+1841 & TXS 1146+189 & 1.256  & 2.572  & 13.057  & 23.243  & -12.110  & -12.042  & 1.170  & 45.731   \\ 
         J1223.8+8039 & S5 1221+80 & 0.473  & 2.291  & 12.372  & 21.687  & -11.440  & -11.507  & 0.858  & 44.883   \\ 
         J1224.9+4334 & B3 1222+438 & 1.075  & 2.467  & 12.695  & 22.984  & -12.127  & -11.576  & 3.562  & 45.738   \\ 
         J1331.2-1325 & PMN J1331-1326 & 0.250  & 2.609  & 13.218  & 23.644  & -12.130  & -11.782  & 2.231  & 43.927   \\ 
         J1353.3+1434 & OP 186 & 0.808  & 2.278  & 13.460  & 21.630  & -12.031  & -11.764  & 1.849  & 45.188   \\ 
         J1516.9+1934 & PKS 1514+197 & 1.070  & 2.505  & 12.971  & 20.798  & -11.415  & -11.394  & 1.050  & 45.093   \\ 
         J1824.1+5651 & 4C +56.27 & 0.664  & 2.389  & 13.151  & 21.466  & -11.388  & -11.198  & 1.550  & 45.357   \\ 
         J1954.6-1122 & TXS 1951-115 & 0.683  & 2.415  & 13.062  & 21.600  & -11.810  & -11.260  & 3.543  & 45.307   \\ 
         J2010.0+7229 & 4C +72.28 & 0.239  & 2.292  & 12.844  & 21.727  & -11.756  & -11.231  & 3.346  & 44.462   \\ 
         J2012.2-1646 & PMN J2012-1646 & 0.239  & 2.697  & 13.135  & 21.668  & -12.394  & -10.984  & 25.687  & 43.636   \\ 
         J2032.0+1219 & PKS 2029+121 & 1.215  & 2.410  & 12.595  & 21.449  & -11.623  & -11.535  & 1.225  & 45.537   \\ 
         J2200.3+1029 & TXS 2157+102 & 0.770  & 2.501  & 12.395  & 19.229  & -12.256  & -11.299  & 9.063  & 45.199   \\ 
         J2357.4-0152 & PKS 2354-021 & 0.812  & 2.268  & 13.161  & 22.917  & -12.074  & -12.133  & 0.874  & 44.896   \\ \hline
            \end{tabular}
            \tablecomments{Columns (1) is the 4FGL name of sources; Columns (2) is the associated source name; Columns (3) is the redshift; Columns (4) is the $\gamma$ -ray photon spectral index; Columns (5) is the Synchrotron radiation peak frequency; Columns (6) is the Inverse Compton scattering peak frequency; Columns (7) is the Synchrotron radiation peak flow; Columns (8) is the Inverse Compton scattering peak flow; Columns (9) is the Compton dominant parameter; Columns (10) is $\gamma$ -ray luminosity.}
\end{table}

\section{Spectrum Identification }  \label{sec:spectrum identification}
The primary objective of this paper is to identify misclassified FSRQ sources among LBLs. We employed Gaussian fitting to determine the dividing threshold based on the 1$\sigma$ deviation of FSRQ near LBLs. As a result, we identified 25 potentially misclassified sources. Subsequently, we applied spectral authentication methods to assess the effectiveness of our screening process. 

\subsection{ The Spectrum data}

 The Sloan Digital Sky Survey (SDSS) \footnote{\scriptsize{\url{https://skyserver.sdss.org/}}} is a major international astronomy project utilizing two custom-designed 2.5-meter telescopes to create a detailed 3D map of the universe (\citealt{2023ApJS..267...44A}). It has collected spectra from over three million astronomical objects, covering about one-third of the sky. The project aims to precisely map a quarter of the sky, calibrate photometric data, and measure distances for over one million galaxies and quasars to study large-scale cosmic structures.
SDSS offers free public access to its database of over 80 million stars, galaxies, and quasars, along with tools for image acquisition, sky browsing, data queries, and visualization. Users can download spectral data via image tools, run SQL queries for photometric and spectroscopic data, or use CasJobs for batch queries and personal database storage.

The NASA/IPAC Extragalactic Database (NED\footnote{\scriptsize{\url{https://ned.ipac.caltech.edu/}}}) is a collaborative astronomical database managed by NASA and the Infrared Processing and Analysis Center (IPAC) at the California Institute of Technology. It provides extensive data on extragalactic systems and celestial objects beyond the Milky Way. This comprehensive resource includes cross-identifications, precise coordinates, redshifts, and key physical parameters of extragalactic objects, along with a curated collection of references and abstracts relevant to extragalactic astronomy.
NED hosts over 773,000 images from 2MASS, published literature, and digitized sky surveys, serving as a valuable tool for astronomers. The database is updated regularly, with revised versions released online every two to three months, ensuring the information remains current and reliable.

To access specific galaxy information within NED, one can visit its official website and utilize its the built-in search tools to query galaxies by name or coordinates. The intuitive interface supports filtering based on multiple parameters to refine search results. After submitting a query, a list of matching galaxies is displayed, allowing users to select a specific galaxy to view comprehensive details, including  celestial coordinates, redshift, luminosity, distance, morphology, and other key astronomical data.
For more advanced searches, NED offers enhanced query capabilities, enabling users to apply multiple filters to locate precise galaxy data.
Our spectral data are primarily sourced from SDSS and NED. Initially, the right ascension and declination of the sources are retrieved from NED by querying the source name. These coordinates are then used to search the SDSS database for the corresponding spectral FITS files, which can be downloaded for further analysis.

Among the 25 potentially misclassified sources identified in our analysis, spectral data were available for 13 sources, while 12 lacked spectral data. Of the 12 sources without spectra:
MG1 J114208+1547, PKS 0346-163, PKS 0437-454, PKS 2029+121, PMN J2012-1646, S5 1221+80,TXS 0404+075, TXS 2157+102, TXS  1951-115, was compiled as BL Lacs in the 3FGL catalog. PMN J1331-1326 is a candidate source for BL Lacs, TXS 1951-115 is classified as FSRQs, and PKS 2029+121 has no clear classification (\citealt{2015ApJS..218...23A}). \cite{2006A&A...455..773V} compiled PKS 0346-163 in the BL Lacs directory \footnote{\scriptsize{\url{https://cdsarc.cds.unistra.fr/viz-bin/nph-Cat/txt?VII/248/bllac.dat.gz}}}. Further spectrum observations of these sources are needed to confirm whether they are misjudged.

\subsection{Qsofitmore}
QSOFITMORE\footnote{\scriptsize{\url{https://github.com/rudolffu/qsofitmore}}} is a Python package designed  for  UV-optical QSO spectra fitting. This package, developed based on PyQSOFit \footnote{\scriptsize{\url{https://github.com/legolason/PyQSOFit}}} (v1.1)(\citealt{2018ascl.soft09008G}; \citealt{2019ApJS..241...34S}), includes additional features tailored for the LAMOST quasar survey (\citealt{2023ApJS..265...25J}) and the survey of quasars behind the Galactic plane (\citealt{2022ApJS..261...32F}). It is now a standalone package released under the same GNU license as.

The spectral fitting process involves several meticulous steps to ensure accurate extraction of quasar properties and robust uncertainty estimation. First, the data are imported, with QSOFITMORE supporting spectra in non-SDSS formats. Galactic extinction is corrected using the Planck 14 dust map combined with the extinction law from \cite{2019ApJ...877..116W}. Each spectrum is then shifted to the rest frame using known redshift values, aligning the data for further analysis. For quasars with low redshifts ($z<1.16$), a Principal Component Analysis (PCA) decomposition separates the host galaxy and quasar components, facilitating the elimination of host galaxy contributions. This step utilizes  five eigenspectra for the host galaxy and 20 for the quasar, producing a pure quasar spectrum for subsequent fitting. We modeled the continuum and pseudo-continuum using multiple components: a power-law function represents the general shape of the quasar spectrum, third-order polynomial accounts for the pseudo-continuum, and a Fe II template models Fe II emission. For quasars with redshifts $ \geq 2.5$, the polynomial component mitigates interference from broad absorption lines. After fitting the continuum, we modeled the emission lines using Gaussian profiles, which fit narrow-line and broad-line components. We fitted individually key emission lines such as $H\alpha$, $H\beta$, Mg II, and for higher redshifts, C III and C IV. The Monte Carlo (MC) run over 50 iterations to assess uncertainties. Each iteration perturbs the fit with random noise from a Gaussian distribution, and the standard deviation of these trials represents the uncertainty in each measured parameter. This comprehensive process ensures the precise decomposition of spectral components and reliable measurements of quasar properties, accounting for both instrumental and intrinsic variations. Table \ref{tab:5} presents the final fitting results. 

\setlength{\tabcolsep}{3pt} 
\begin{sidewaystable}[ht]
    \caption{Spectral fitting results}
    \label{tab:5}
    \centering
    \begin{tabular}{cccccccccccc}
    \hline \hline
        sourcename & redshift & MgII ew  & MgII ew\_err & H$\beta$ ew & H$\beta$ ew\_err & H$\alpha$ ew & H$\alpha$ ew\_err & C III ew & C III ew\_err & CIV ew  &CIV ew\_err  \\ 
        \normalsize(1) & \normalsize(2) & \normalsize(3) &\normalsize(4) & \normalsize(5)   &\normalsize(6) &\normalsize(7) &\normalsize(8) &\normalsize(9) &\normalsize(10) &\normalsize(11)&\normalsize(12)\\
        \hline\hline
            OJ 448 & 0.548  & 1.603  & 0.521  & 1.105  & 0.516  & 0.000  & 0.156  & ~ & ~ & ~ &   \\ 
        OP 186 & 0.808  & 2.992  & 0.769  & 6.307  & 1.595  & ~ & ~ & ~ & ~ & ~ &   \\ 
        7C 1032+4424 & 0.444  & 8.444  & 5.023  & 0.000  & 0.939  & 2.647  & 4.510  & ~ & ~ & ~ &   \\ 
        MG1 J114208+1547 & 0.299  & 20.052  & 4.448  & 0.000  & 0.434  & 12.865  & 2.648  & ~ & ~ & ~ &   \\ 
        GB6 J0941+2721 & 0.631  & 31.959  & 7.914  & 7.747  & 3.487  & ~ & ~ & ~ & ~ & ~ &   \\ 
        PKS 2354-021 & 0.812  & 3.501  & 0.538  & 2.109  & 1.200  & ~ & ~ & ~ & ~ & ~ &   \\ 
        B3 1222+438 & 1.075  & 1.797  & 2.101  & 0.000  & 0.000  & ~ & ~ & 7.307  & 4.161  & ~ &   \\ 
        PKS 1514+197 & 1.070  & 1.261  & 0.450  & 17.344  & 2.964  & ~ & ~ & 3.355  & 1.629  & ~ &   \\ 
        S4 0707+47 & 1.292  & 4.649  & 0.926  & ~ & ~ & ~ & ~ & ~ & ~ & ~ &   \\ 
        GB6 J0934+3926 & 0.236  & ~ & ~ & 0.000  & 1.572  & 114.409  & 89.021  & ~ & ~ & ~ &   \\ 
        TXS 1146+189 & 1.256  & 1.333  & 1.194  & ~ & ~ & ~ & ~ & 2.108  & 0.959  & ~ &   \\ 
        4C+56.27 & 0.664  & 1.109  & 0.183  & ~ & ~ & ~ & ~ & ~ & ~ & ~ &   \\ 
        NVSS J112903+375655 & 4.090  & ~ & ~ & ~ & ~ & ~ & ~ & 2.682  & 0.887  & 1.059  & 0.611   \\ \hline
    \end{tabular}
    \tablecomments{Columns (1) is the name of sources; Columns (2) is the redshift; Columns (3) is MgII equivalent width of emission lines; Columns (4) is MgII equivalent width  err of emission lines; Columns (5) H$\beta$ equivalent width of emission line; Columns (6) is H$\beta$ equivalent width err of emission line; Columns (7) is H$\alpha$ equivalent width of emission line; Columns (8) is H$\alpha$ equivalent width  err of emission line; Columns (9) is C III equivalent width of emission line; Columns (10) is C III equivalent width err of emission line; Columns (11) is CIV equivalent width err of emission line; Columns (12) is CIV equivalent width err of emission line. The units of the equivalent width and error of all emission lines are $\mathring{\mathrm{A}}$. The blank space indicates that there is no wide emission line component of this type.}
\end{sidewaystable}

\subsection{The sources with $ EW< 5\mathring{\mathrm{A}}$}
\begin{figure*}[htbp]
	\centering
	\includegraphics[width=0.8\linewidth]{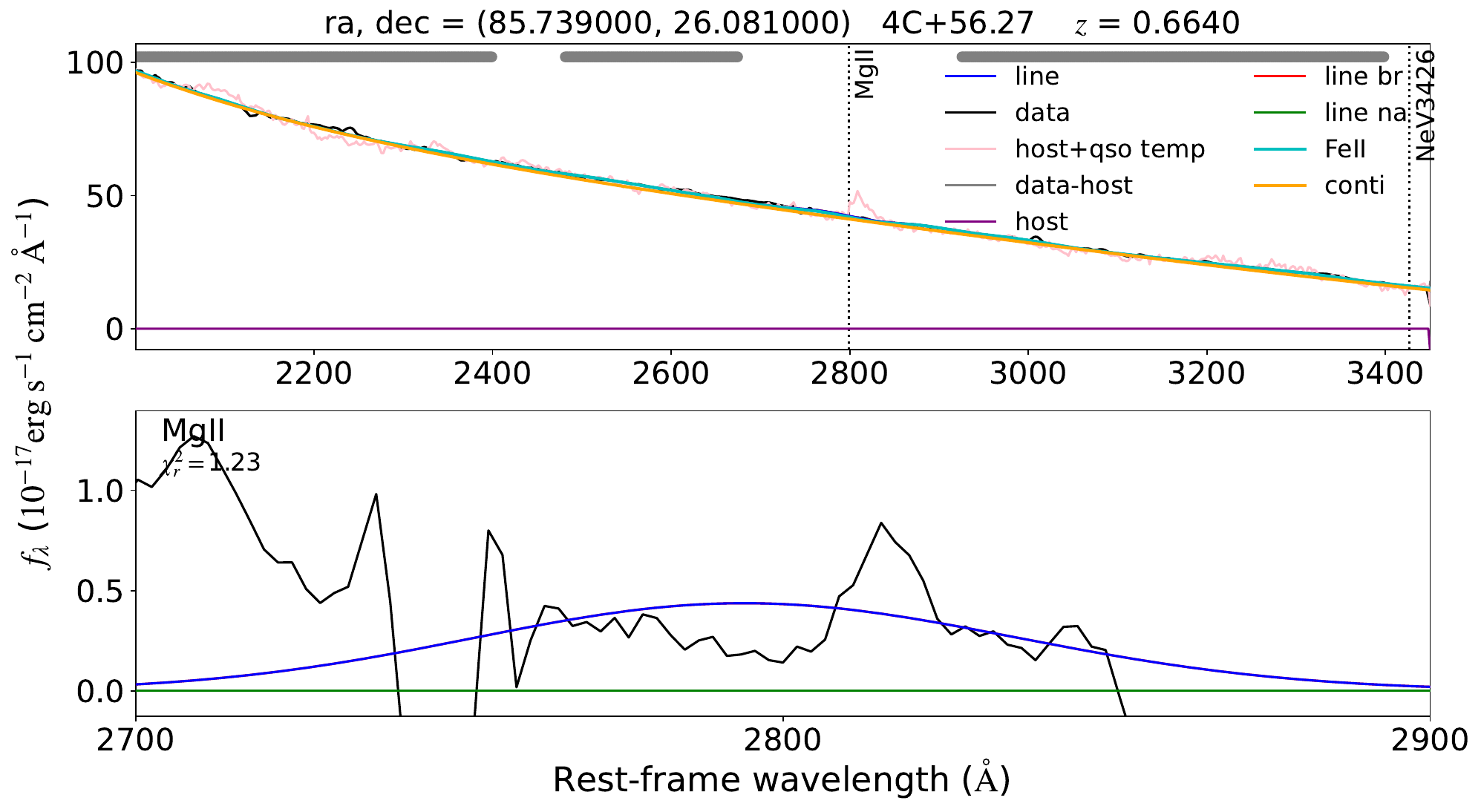}
	\caption{The spectral fitting of quasar 4C+56.27 at redshift \( z = 0.6640 \), observed at coordinates (RA, Dec) = (85.739000, 26.081000). In the upper panel, the black line represents the total dereddened spectrum, while yellow represents the continuum component. Cyan denotes the Fe II emission templates, blue represents the combined emission lines, red indicates broad-line components, and green indicates narrow-line components. The purple line represents the host galaxy contribution, the gray line represents the quasar component, and the pink line denotes the spectrum reconstructed from the host and quasar components via PCA. In the lower panel, the Mg II line at 2800 $\mathring{\mathrm{A}}$ is highlighted with its equivalent width (EW) labeled, showing details of the emission line fitting.}
	\label{fig:7}
\end{figure*}

4FGL J1824.1+5651: Figure \ref{fig:7} shows strong, broad emission lines in the spectrum of 4C+56.27. The equivalent width of its emission lines is less than $5 \mathring{\mathrm{A}}$, and it has been considered a BL Lac object by some authors due to nearly featureless spectrum (\citealt{1986AJ.....91..494L}), and high, variable polarization (\citealt{1982AJ.....87..859P}; \citealt{1985ApJS...59..513A}). \cite{1986AJ.....91..494L} determined a tentative redshift of z=0.664.

\begin{figure*}[htbp]
	\centering
	\includegraphics[width=0.8\linewidth]{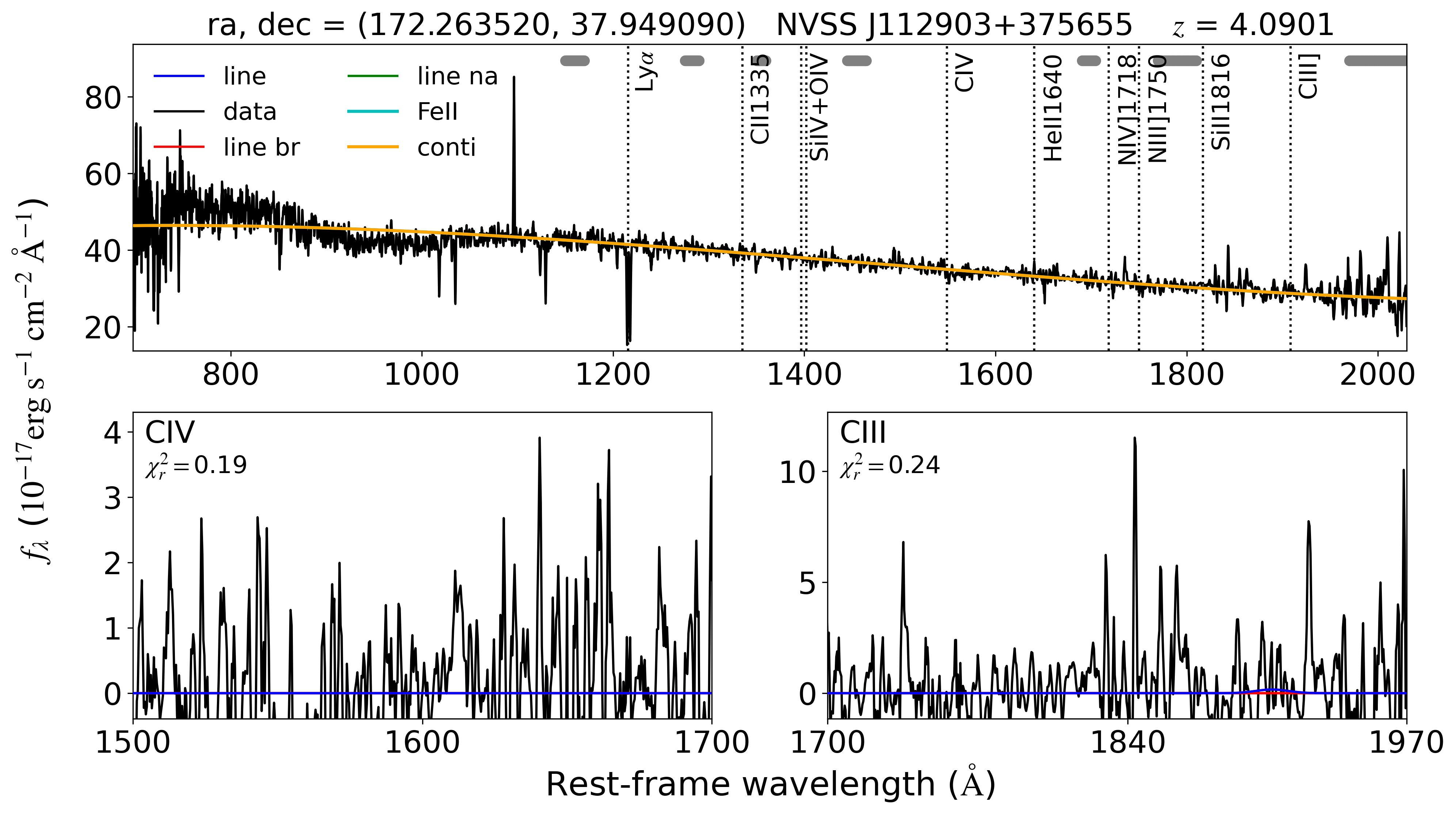}
	\caption{The spectrum of NVSS J112903+375655. This image represents the spectral analysis of the quasar NVSS J112903+375655, located at coordinates (ra, dec)=(172.263520, 37.949090) with a redshift of z=4.0901. The plot highlights the Lyman-alpha (Lya) and CIV emission lines, with data points and fitted lines indicated. The spectral features are overlaid on a continuum model, with line broadening and other details marked. The flux density is denoted by $f_{1}(10^{-17} ergs^{-1}cm^{-2}A^{-1})$, and the rest-frame wavelength is plotted on the x-axis in Angstroms. The specific values of $\chi^2=0.19$ and $\chi^2=0.24$ likely refer to parameters related to the Gaussian fitting of the emission lines. Key emission lines such as CIV and C III are labeled, and the plot captures the detailed structure of the quasar's spectrum, essential for understanding its physical properties and redshift.
}
	\label{fig:8}
\end{figure*}

4FGL J1128.8+3757: NVSS J112903+375655 was included in the Directory of Radio Sources in 2002 (\citealt{2002ApJS..143....1M}), and classified as QSQ (\citealt{2009ApJS..180...67R}; \citealt{2009MNRAS.396..223D}). Only very weak emission lines are present in the spectrum of this source, as showen in Figure \ref{fig:8}. The lower limit of the equivalent width calculated by \cite{2017ApJ...851..135P}, ranges from $0.2\mathring{\mathrm{A}}$  to $0.5\mathring{\mathrm{A}}$, which is consistent with the results of our spectral fitting.

\begin{figure*}[htbp]
	\centering
	\includegraphics[width=0.8\linewidth]{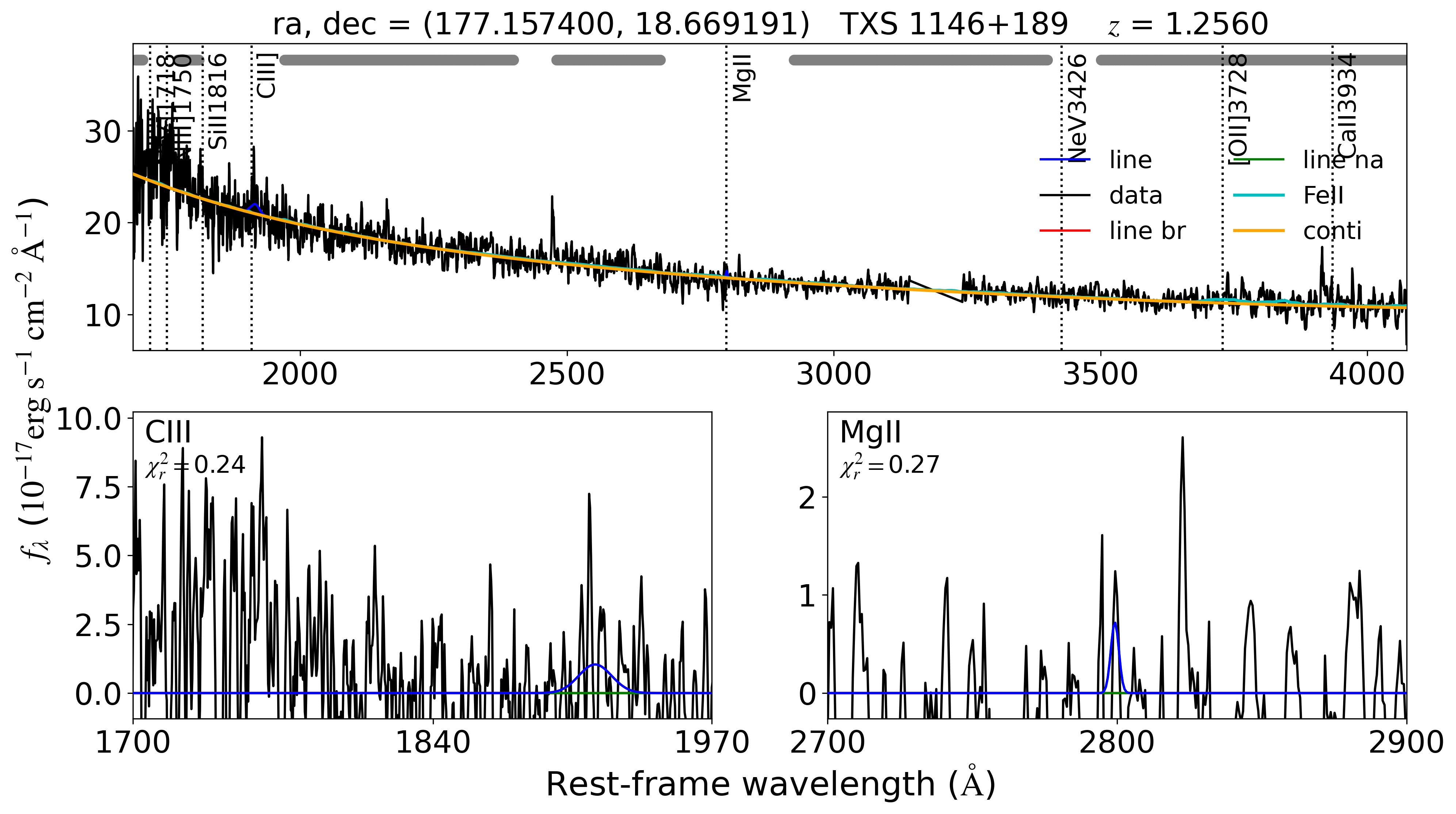}
	\caption{The spectrum of TXS 1146+189. In this figure, all the wide emission lines $ EW< 5\mathring{\mathrm{A}}$.}
	\label{fig:9}
\end{figure*}

4FGL J1148.6+1841: TXS 1146+189 is a Flat-Spectrum Radio Source (\citealt{2007ApJS..171...61H}), we did not find strong, broad emission lines in its spectrum during the fitting process, as shown in Figure \ref{fig:9}. Whether it is a misidentified source requires further spectroscopic observations.

\begin{figure*}[htbp]
	\centering
	\includegraphics[width=0.8\linewidth]{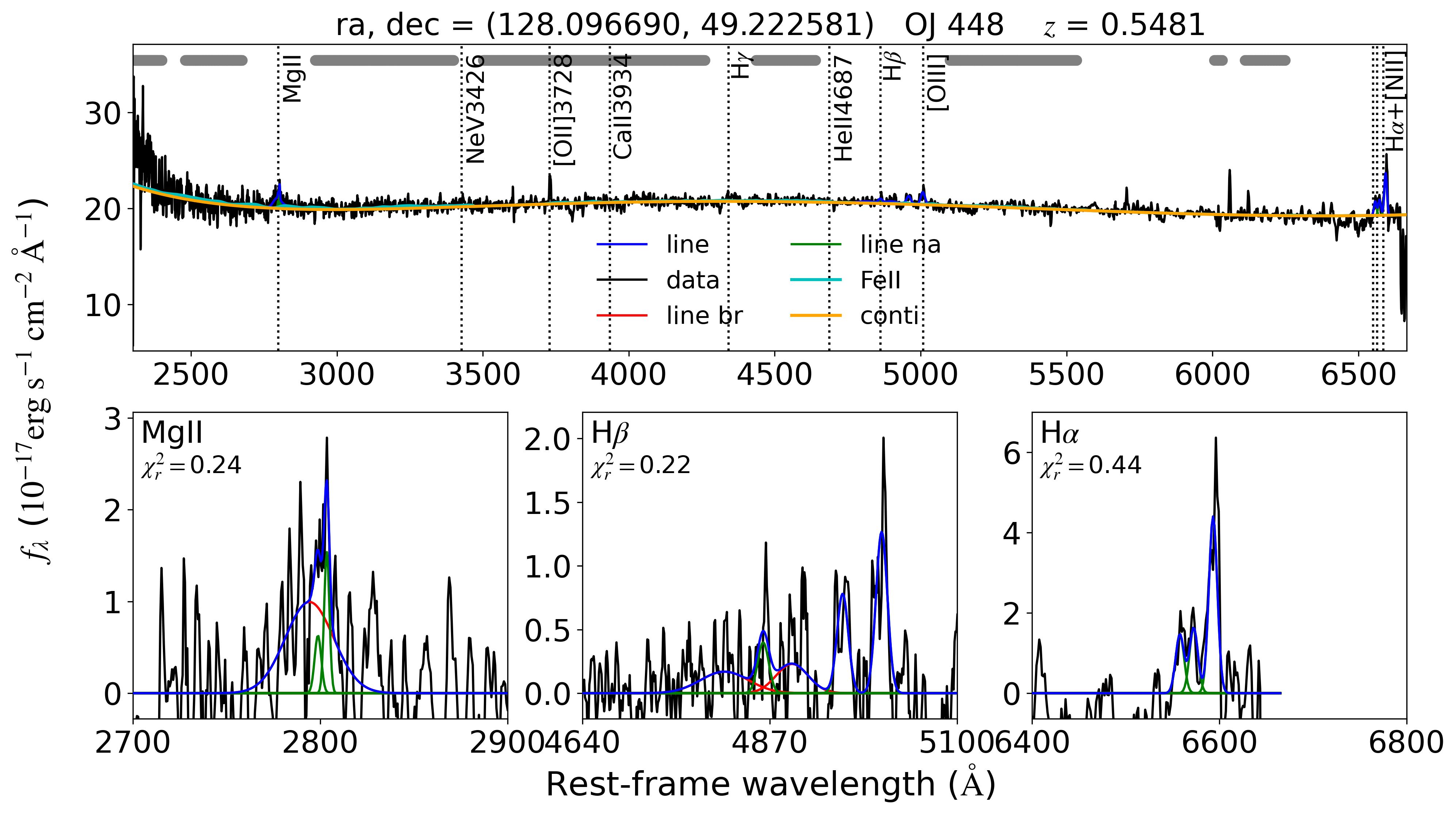}
	\caption{Example of spectral fitting for the quasar OJ 448 at \( z = 0.5481 \). Black lines denote the total dereddened spectrum, yellow lines denote the continuum (modeled by a power-law function and a third-order polynomial), cyan lines denote the Fe II emission template, blue lines denote the entire emission line profile, red lines denote the broad-line components, and green lines denote the narrow-line components.In the figure below, we can see the obvious broad emission line composition of MgII, and the EW of its emission line is calculated to be 1.603$\mathring{A}$ by fitting.}
	\label{fig:10}
\end{figure*}

4FGL J0832.4+4912: OJ 448  was classified as a BL Lac in the 3FGL Catalog. We found strong and broad emission lines in its spectrum during the fitting process, as shown in Figure \ref{fig:10}, but its MgII equivalent width is $1.603\mathring{A}$.

\begin{figure*}[htbp]
	\centering
	\includegraphics[width=0.8\linewidth]{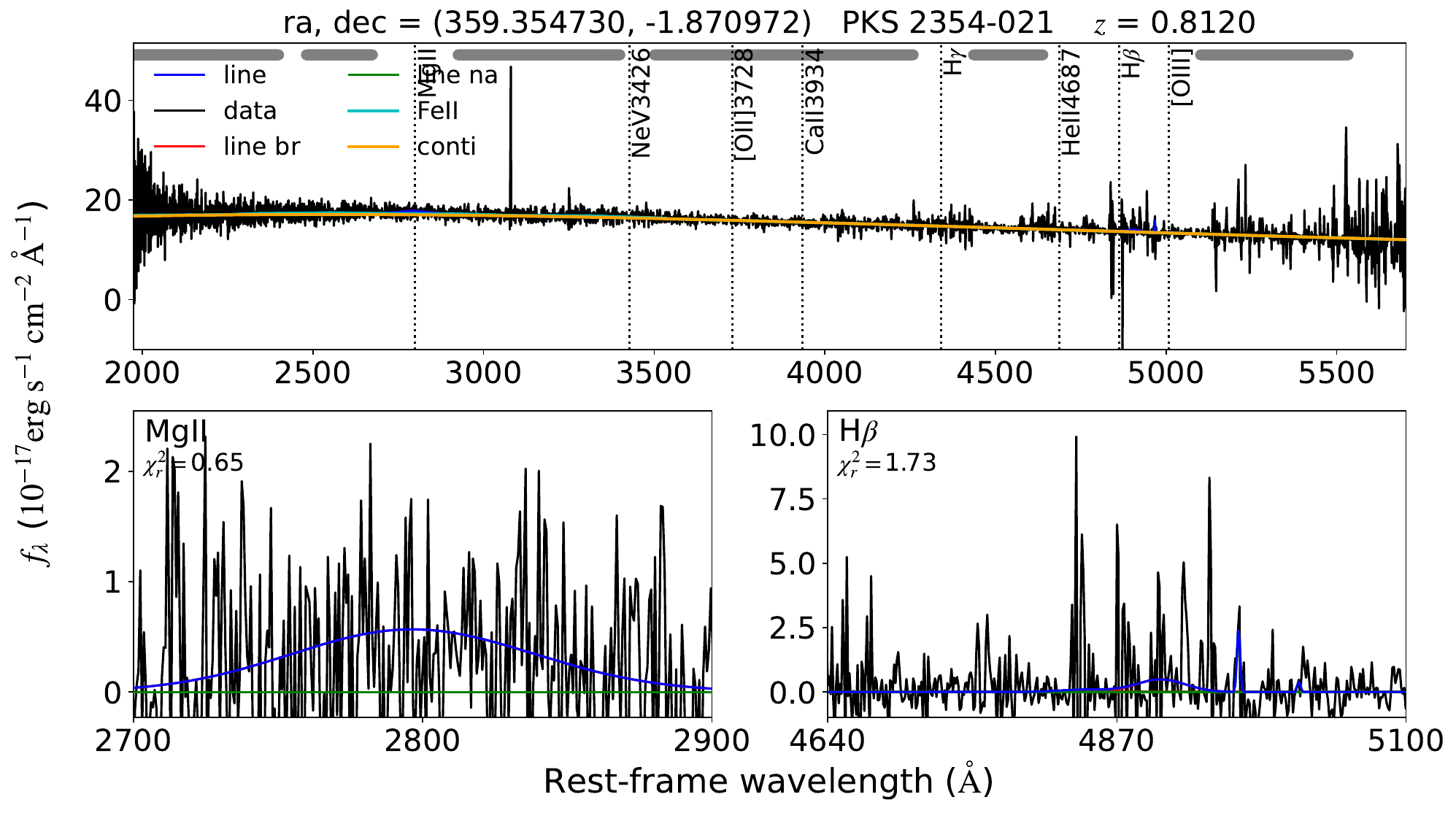}
	\caption{The spectrum of PKS 2354-021. Example of spectral fitting for the quasar PKS 2354-021 at \( z = 0.8120 \). Black lines denote the total dereddened spectrum, yellow lines denote the continuum (modeled by a power-law function and a third-order polynomial), cyan lines denote the Fe II emission template, blue lines denote the entire emission line profile, red lines denote the broad-line components, and green lines denote the narrow-line components. In the figure below, we can see the obvious broad emission line composition of MgII, and the EW of its emission line is calculated to be 3.501$\mathring{A}$ by fitting.}
	\label{fig:11}
\end{figure*}

4FGL J2357.4-0152: PKS 2354-021 was classified as a BL Lac (\citealt{2006A&A...455..773V}). In this article,  we argue that PKS 2354-021 is a questionable BL Lac. We found strong and broad emission lines in its spectrum during the fitting process, as shown in Figure \ref{fig:11}, but its equivalent widths are 3.501$\mathring{A}$ for MgII and 2.109$\mathring{A}$ for $H\beta $.

\subsection{The sources with $ EW \geq5\mathring{\mathrm{A}}$}
\subsubsection{OP 186}
\begin{figure*}[htbp]
	\centering
	\includegraphics[width=0.8\linewidth]{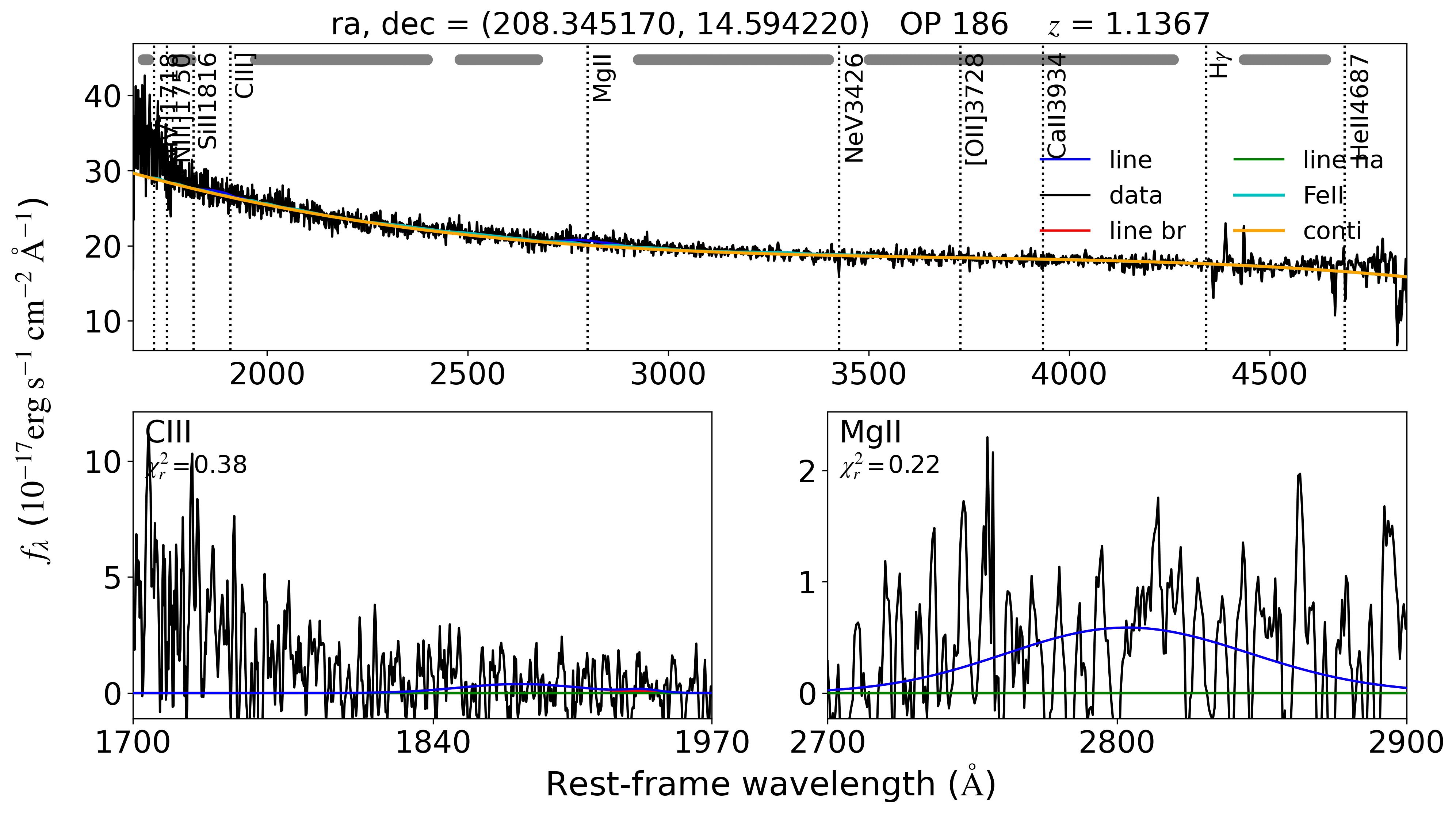}
	\caption{Spectral fitting for the quasar OP 186 at redshift z=1.1367. The black curve shows the observed dereddened spectrum. Yellow represents the continuum model (fpl + fpoly ), cyan denotes the Fe II templates ($f_{Fe II}$ ), blue corresponds to the total emission line fit, red represents broad-line components, and green represents narrow-line components. Insets highlight detailed fits for MgII ( $ \chi^2=0.85$) and $H\beta$ ($ \chi^2=5.36$), demonstrating line decomposition accuracy.}
	\label{fig:12}
\end{figure*}

4FGL J1353.3+1434: It was classified as a BL Lac in the 3FGL Catalog (\citealt{2015ApJS..218...23A}). The study of OP 186 was carried out on the assumption that it is a BL Lacs (\citealt{2013ApJ...764..135S}; \citealt{2014A&A...572A..59M}; \citealt{2022MNRAS.515.4810L}). OP 186 may be a FSRQ with a CD of 1.849. We found strong and broad emission lines in its spectrum during the fitting process, as shown in Figure \ref{fig:12}, and its $H\beta$ equivalent width  $6.307\mathring{\mathrm{A}}$.

\subsubsection{GB6 J0941+2721}
GB6 J0941+2721 was a QSO in the past (\citealt{2009MNRAS.396..223D}). Despite being classified as a BL Lac object in the 4FGL catalog, we found that GB6 J0941+2721 has strong and broad emission lines by fitting its SDSS spectrum. The equivalent width of its $H\alpha$ emission line is 114.109 $ \mathring{\mathrm{A}}$.

\begin{figure*}[htbp]
	\centering
	\includegraphics[width=0.8\linewidth]{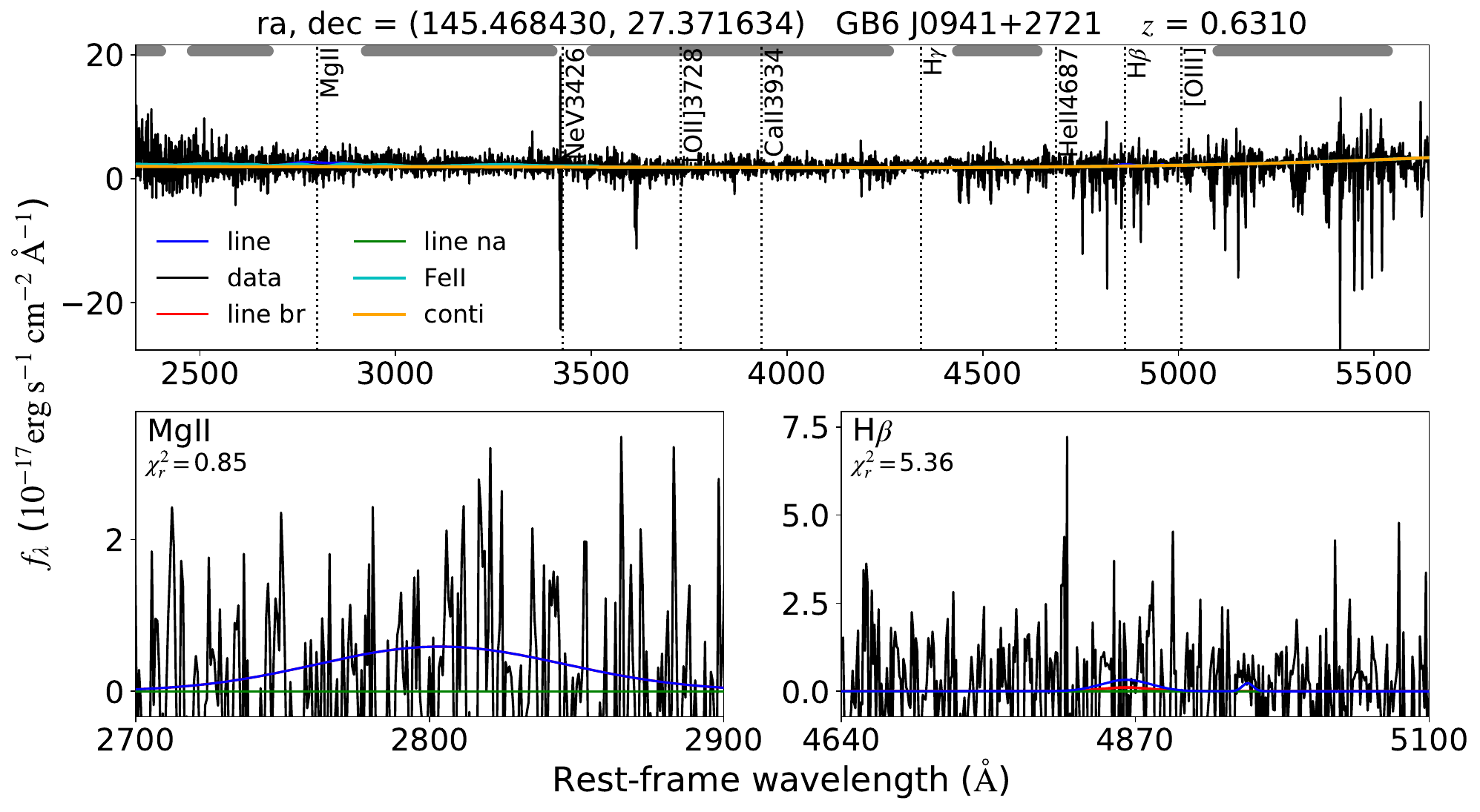}
	\caption{Spectral fitting for the quasar GB6 J0941+2721 at redshift z=0.6310. The black curve shows the observed dereddened spectrum. Yellow represents the continuum model (fpl + fpoly ), cyan denotes the Fe II templates ($f_{Fe II}$ ), blue corresponds to the total emission line fit, red represents broad-line components, and green represents narrow-line components. Insets highlight detailed fits for MgII ( $ \chi^2=0.85$) and $H\beta$ ($ \chi^2=5.36$), demonstrating line decomposition accuracy.}
	\label{fig:13}
\end{figure*}

\subsubsection{B3 1222+438}

\begin{figure*}[htbp]
	\centering
	\includegraphics[width=0.8\linewidth]{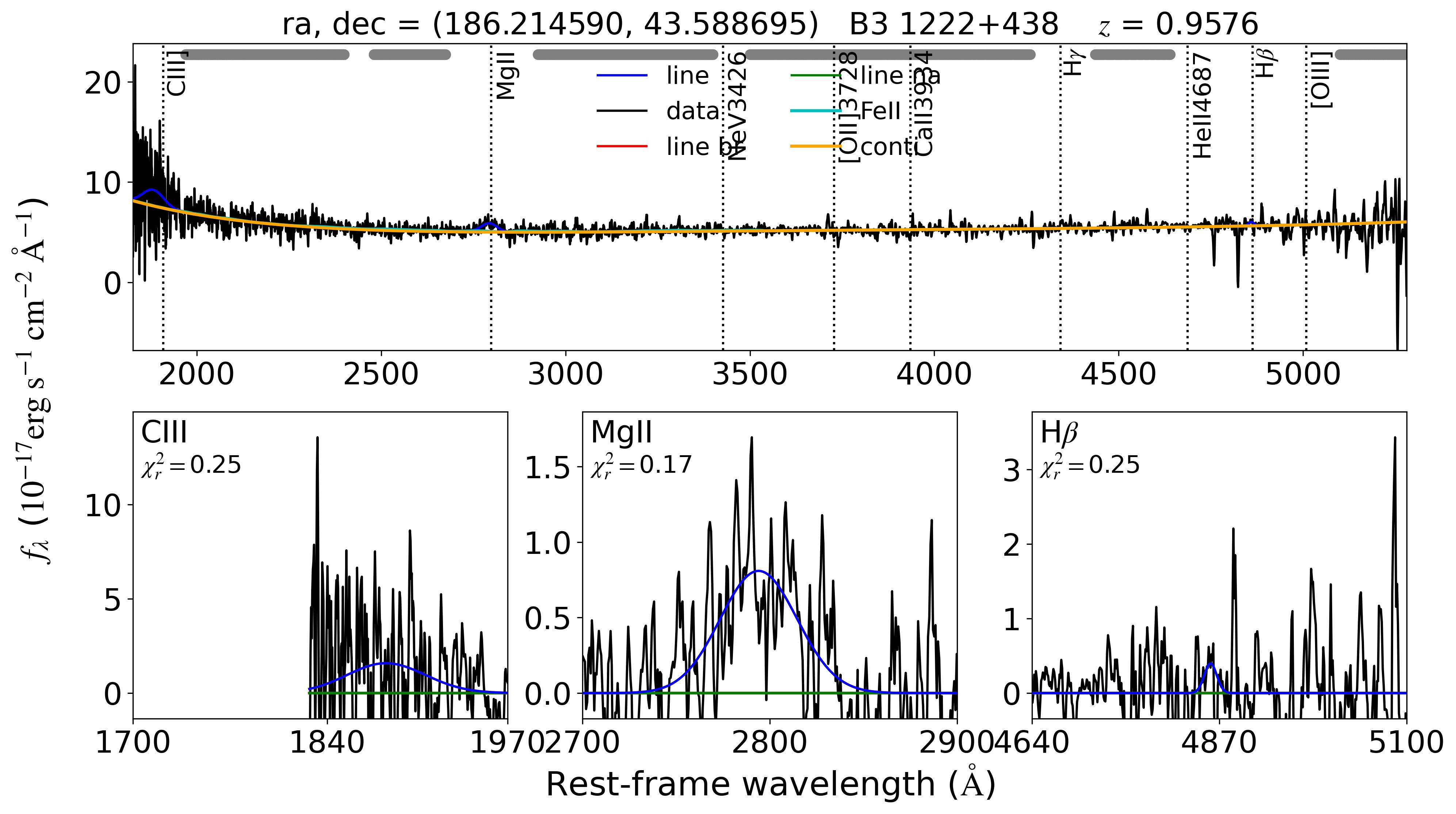}
	\caption{Spectral fitting of quasar B3 1222+438 at redshift \( z = 0.9576 \), observed at coordinates (ra, dec) = (186.214590, 43.588695). The top panel shows the full dereddened spectrum with black representing the observed data. Yellow represents the continuum component, cyan denotes the Fe II emission, blue represents the total emission line profiles, and red highlights specific line fits. Key emission lines such as C III, Mg II, H\(\beta\), and O III are labeled. The lower panels provide zoomed-in views of individual line fits for C III (left), Mg II (center), and H\(\beta\) (right), with blue lines representing the model fits. The reduced chi-square (\( \chi_r^2 \)) values for each line fitting are displayed, indicating the fit quality for each component.}
	\label{fig:14}
\end{figure*}

4FGL J1224.9+4334: B3 1222+438 has been considered an FSRQ in the research of Sowards-Emmerd and David's  (\citealt{2003ApJ...590..109S}). However, in 2003, B3 1222+438 was designated a BL Lac because its emission line equivalent width was less than 5 $\mathring{\mathrm{A}}$ (\citealt{2008AJ....135.2453P}). A list of physical parameters for these sources is also provided\footnote{\scriptsize{\url{https://content.cld.iop.org/journals/1538-3881/135/6/2453/revision1/aj268053_mrt5.txt}}}\footnote{\scriptsize{\url{https://content.cld.iop.org/journals/1538-3881/135/6/2453/revision1/aj268053_mrt6.txt}}}. It was classified as a BL Lac in the 3FGL catalog. We found a broad emission line in its spectrum, as shown in Figure \ref{fig:14}. The equivalent width of its C III emission line is 7.307 $ \mathring{\mathrm{A}}$.

\subsubsection{PKS 1514+197}
\begin{figure*}[htbp]
	\centering
	\includegraphics[width=0.8\linewidth]{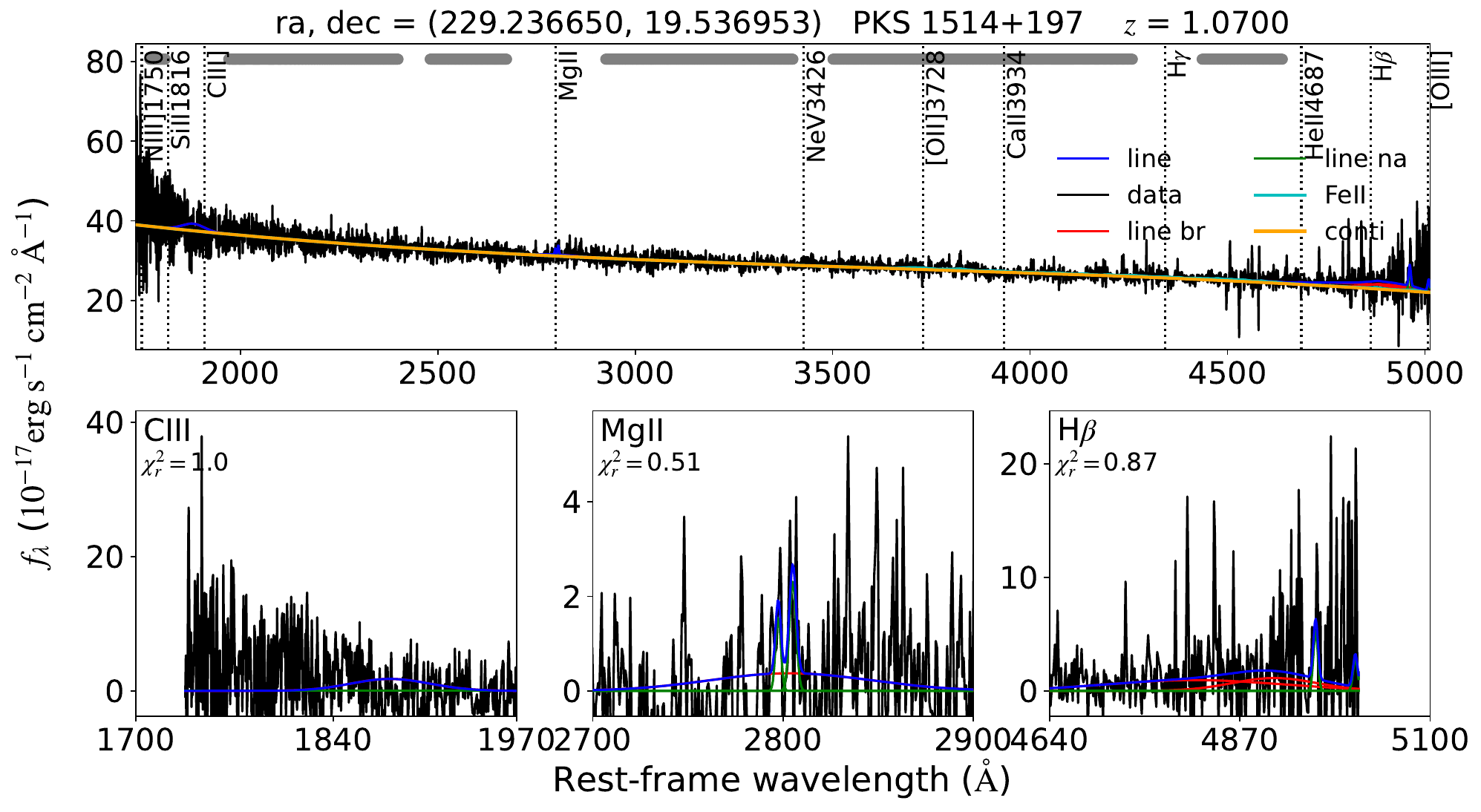}
	\caption{Spectral fitting for the quasar PKS 1514+197 at redshift z=1.070. The black curve represents the observed dereddened spectrum. Yellow denotes the continuum model (fpl + fpoly ), cyan represents the Fe II templates ($f_{Fe II}$ ), blue corresponds to all fitted emission lines, red shows the broad-line components, and green shows the narrow-line components. Insets illustrate detailed fits for specific lines: C III, MgII, and $H\beta$, with reduced chi-squared ($\chi_r^2$) values indicating goodness of fit.}
	\label{fig:15}
\end{figure*}

PKS 1514+197 was compiled in the BL Lac directory as early as 2006 (\citealt{2006A&A...455..773V}). During its spectral fitting in Figure \ref{fig:15}, we found a $H\beta$ broad emission line component, and we calculated that its emission line equivalent width is 17.344  $ \mathring{\mathrm{A}}$. Therefore, our screening of this source is effective.

\subsubsection{S4 0707+47}

S4 0707+47 was classified as a QSO (\citealt{2006A&A...455..773V}), in their article, S4 0707+47 does not have a more specific classification and was classified as a BL Lacs in 3FGL catalog.  Our result is consistent with the findings of \citealt{2021Univ....7..372F}, who also reclassified it as an FSRQ. In its spectrum, we found that it has a strong Mg II broad emission line, and we calculated its EW, which is greater than 5 $\mathring{\mathrm{A}}$  within the error margin, as shown in Table \ref{tab:5}.
\begin{figure*}[htbp]
	\centering
	\includegraphics[width=0.8\linewidth]{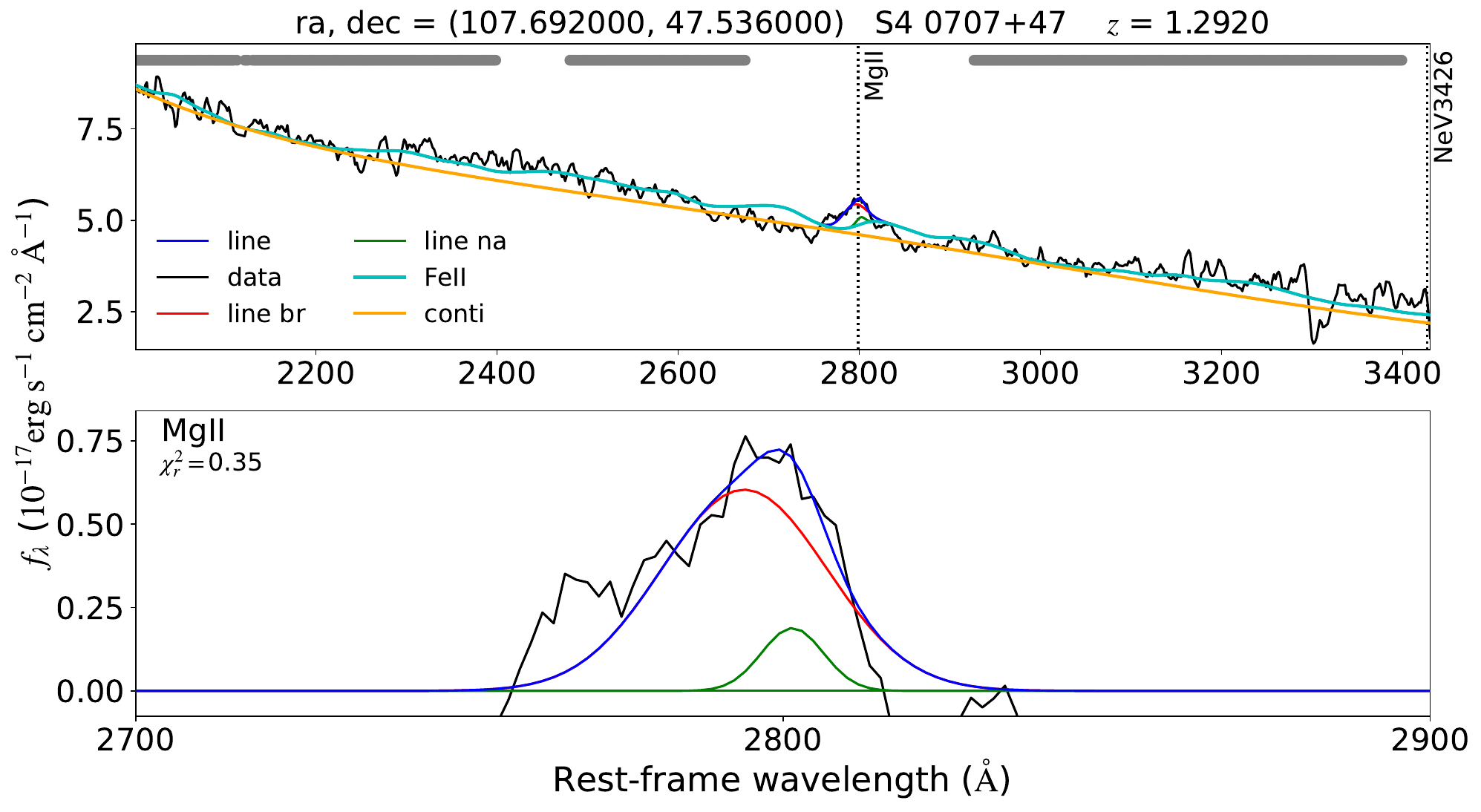}
	\caption{The spectrum of S4 0707+47.  This figure shows examples of spectral fitting for a low-redshift quasar observed by LJT (a) and a high-redshift quasar observed by P200 (b). Black lines represent the total dereddened spectra, yellow lines represent the continua (fpl + fpoly), cyan lines represent the Fe II templates ($f_{Fe II}$), blue lines represent the overall emission lines, red lines represent the broad-line components, and green lines represent the narrow-line components. For a spectrum that is successfully decomposed into a host-galaxy component and a pure quasar component, the purple line represents the host component, the gray line represents the pure quasar component, and the pink line represents the spectrum reconstructed from the PCA host and quasar components.
}
	\label{fig:16}
\end{figure*}

\subsubsection{GB6 J0934+3926}
\begin{figure*}[htbp]
	\centering
	\includegraphics[width=0.8\linewidth]{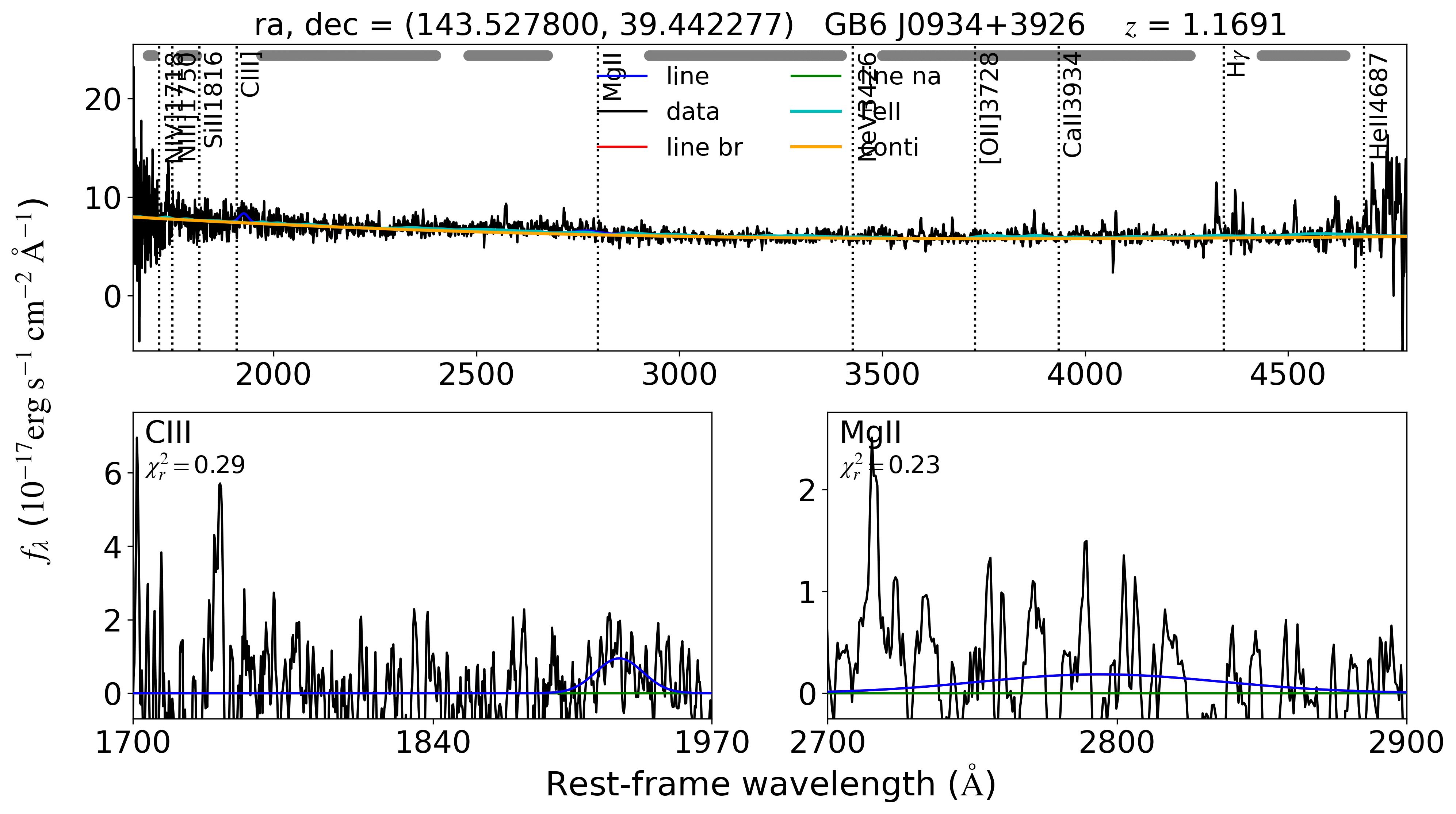}
	\caption{The spectrum of GB6 J0934+3926. This image illustrates the spectral fitting results for the celestial object GB6 J0934+3926, with coordinates of right ascension 143.527800° and declination 39.442277°, and a redshift value of z=1.1691. The plot displays the continuum and emission lines of the spectrum, including C III and MgII. The fitting process involves correction for galactic extinction, redshift correction, Principal Component Analysis (PCA) decomposition, modeling of the continuum and pseudo-continuum, and Gaussian fitting of emission lines. The curves and legends in the plot represent different steps and components of the fitting process, where $f_{2}(10^{-17}\text{erg}\,s^{-1}\,cm^{-2}\,A^{-1})$ denotes the flux density of the spectrum.
}
	\label{fig:17}
\end{figure*}

4FGL J0934.3+3926: GB6 J0934+3926 is classified as an AGN (\citealt{2006A&A...455..773V}), but no more specific classification is available. It was also classified as a BL Lac in the 3FGL catalog. Based on two features: (1) no emission line with a measured rest-equivalent width greater than 5 $\mathring{\mathrm{A}}$, and (2) no measured Ca II H/K depression greater than 40\%, GB6 J0934+3926 was classified as a BL Lac (\citealt{2008AJ....135.2453P}). We found a broad emission line in its spectrum, as shown in Figure \ref{fig:17}.

\subsubsection{7C 1032+4424}
4FGL J1035.6+4409: The spectrum of 7C 1032+4424 contains three broad emission line components: MgII, $H_\beta$, and $H_\alpha$. Due to its low redshift ($z<1.16$), we applied the principal component analysis (PCA) decomposition method to separate the host galaxy and quasar components, thereby eliminating the host galaxy contribution. The final result is the MgII emission line equivalent width ($EW=8.444\mathring{\mathrm{A}}$).
\begin{figure*}[htbp]
	\centering
	\includegraphics[width=0.8\linewidth]{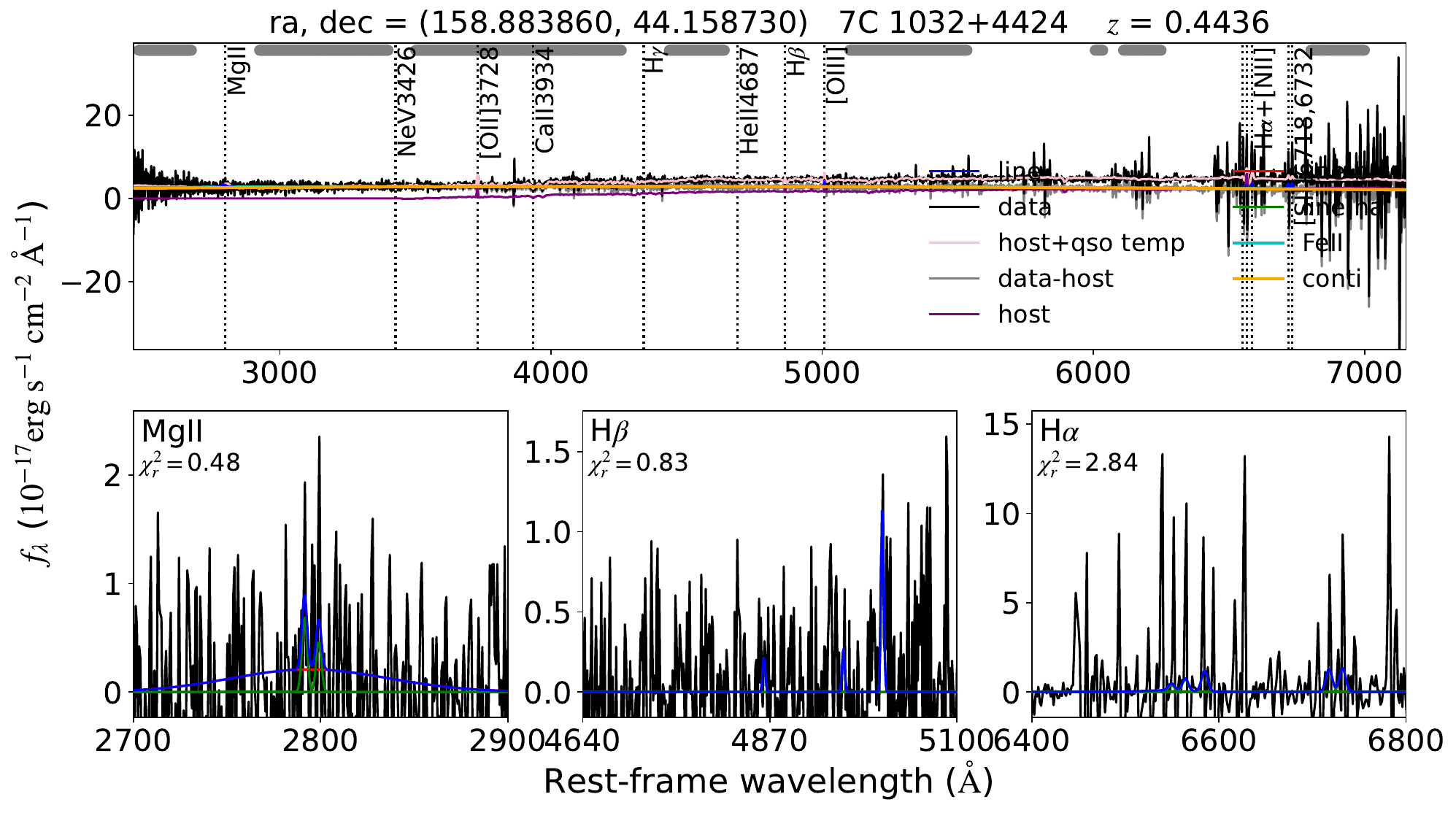}
	\caption{The spectrum of 7C 1032+4424. This image illustrates the spectral fitting results for the celestial object 7C 1032+4424, with coordinates of right ascension 158.883860° and declination 44.158730°, and a redshift value of z=0.4436. The plot displays the continuum and emission lines of the spectrum, including MgII, $H_\beta$ and $H_\alpha$. The fitting process involves correction for galactic extinction, redshift correction, Principal Component Analysis (PCA) decomposition, modeling of the continuum and pseudo-continuum, and Gaussian fitting of emission lines.}
	\label{fig:18}
\end{figure*}

\subsubsection{MG1 J114208+1547}
4FGL J1142.0+1548: MG1 J114208+1547 was classified as a BL Lac in  the 3FGL catalog. By fitting its SDSS spectrum, we found that its Mg II $EW=20.052\mathring{\mathrm{A}}$ and $H_\alpha$ $EW=12.865\mathring{\mathrm{A}}$, with two emission lines of equal width greater than 5$\mathring{\mathrm{A}}$.
\begin{figure*}[htbp]
	\centering
	\includegraphics[width=0.8\linewidth]{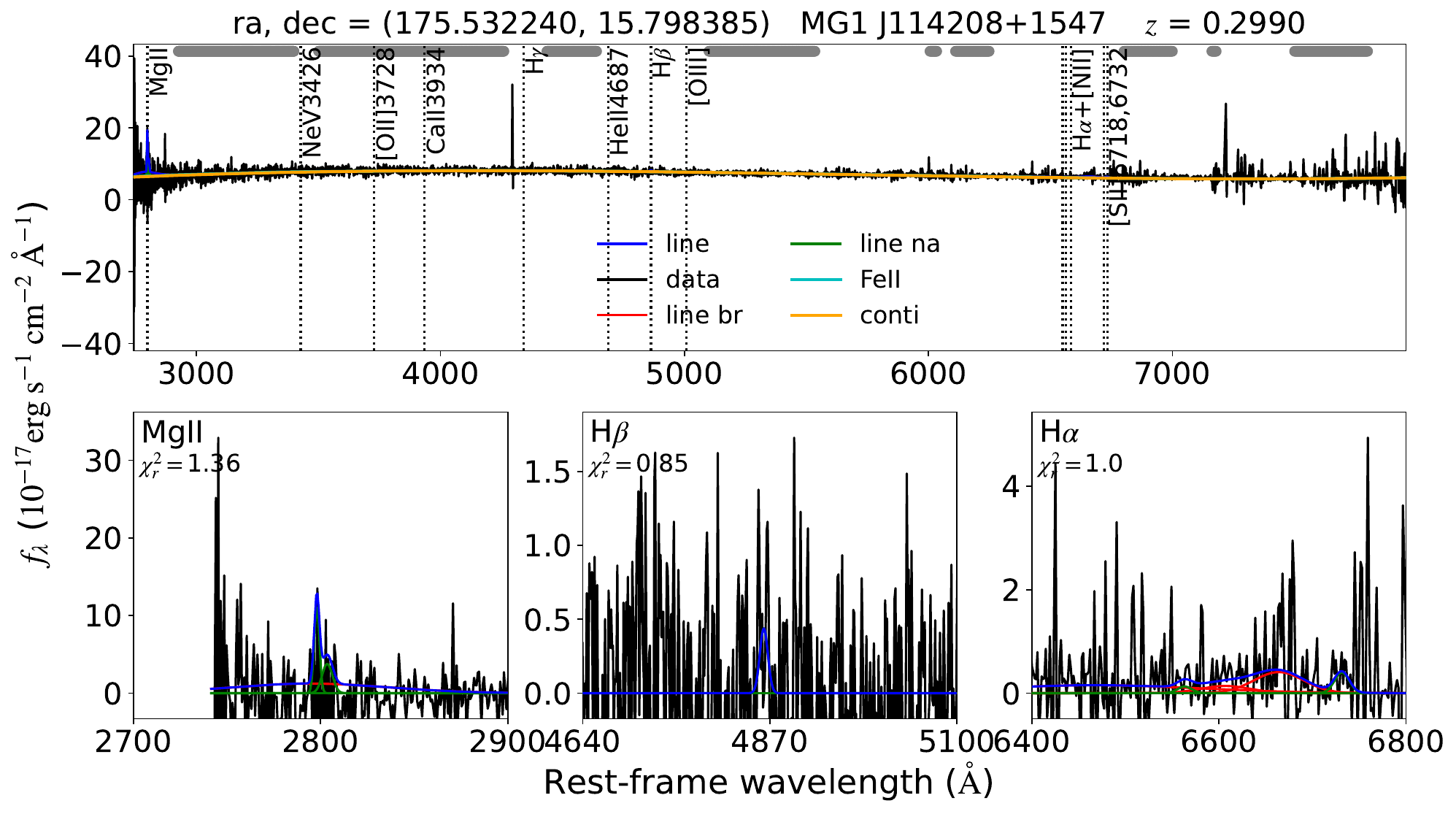}
	\caption{The spectrum of MG1 J114208+1547. This image illustrates the spectral fitting results for the celestial object  MG1 J114208+1547, with coordinates of right ascension 175.532240° and declination 15.798385°, and a redshift value of z=0.299. The plot displays the continuum and emission lines of the spectrum, including MgII, $H_\beta$ and $H_\alpha$. The fitting process involves correction for galactic extinction, redshift correction, Principal Component Analysis (PCA) decomposition, modeling of the continuum and pseudo-continuum, and Gaussian fitting of emission lines.}
	\label{fig:19}
\end{figure*}

\section{Summary}   \label{sec:summary}

This study utilized the 4LAC-DR2 from Fermi-LAT to curate a sample comprising 312 LBLs and 694 FSRQs. We sourced multiband data for both LBL and FSRQ samples from the SSDC and modeled their energy spectra using a logarithmic parabolic approach. This methodology enabled the precise determination of key physical parameters, including the synchrotron radiation peak frequency (\(\log \nu_{\text{peak}}^{\text{syn}}\)), Compton dominance (CD),  and other relevant metrics. A comparative analysis of the LBL and FSRQ distributions identified 30 LBL sources that met the specific criteria for $\Gamma_\gamma$ , $\log \nu_{\text{peak}}^{\text{syn}}$, and the correlation between $\Gamma_\gamma$ and \(\log L_\gamma\). By applying a stringent final screening condition based on  $CD > 0.776$, 25 FSRQs potentially misidentified as LBLs were uncovered.
Statistical analysis confirmed that the SED of blazars can be effectively modeled using a logarithmic parabola. 
Subsequent analysis indicated that sources such as OP 186, GB6 J0941+2721, B3 1222+438, PKS 1514+197, S4 0707+47, and GB6 J0934+3926, which display broad emission lines in their optical spectra with equivalent widths (EW) exceeding 5$\mathring{\mathrm{A}}$, are likely misclassified as LBLs. In conclusion, this study provides significant insights into the classification and physical properties of LBLs and FSRQs, offering a methodology for distinguishing between these sources based on their energy spectra and other critical parameters. 

\begin{acknowledgments}
We acknowledge the use of datas, analysis tools, and services from the Open Universe platform, the ASI Space Science Data Center (SSDC), the \textit{Qsofitmore} Tools, the Astrophysics Data System (ADS), and the NED. 
 I would like to extend my sincere gratitude to Dr.Dong for her invaluable assistance with spectrum fitting and Dr.Zhou for his help with the fitting of SED. We sincerely thank the anonymous reviewers for their valuable comments and suggestions, which have greatly improved the quality of this manuscript. This work is supported by the National Natural Science Foundation of China (grant Nos.12363002, 12163002). The authors would like to express their gratitude to EditSprings (https://www.editsprings.cn ) for the expert linguistic services provided.
\end{acknowledgments}

\bibliography{reference}{}
\bibliographystyle{aasjournal}

\end{document}